\documentstyle{mn-1.4}

\newif\ifAMStwofonts
\AMStwofontstrue

\def\sun{\ifmmode\odot\else$\odot$\fi}
\def\HI{\hbox{H\,{\sc i}}}
\def\HII{\hbox{H\,{\sc ii}}}

\def\Halpha{\hbox{\rm H$\alpha$}}

\def\kms{km~s$^{-1}$}

\ifoldfss
  \ifCUPmtlplainloaded \else
    \NewTextAlphabet{textbfit} {cmbxti10} {}
    \NewTextAlphabet{textbfss} {cmssbx10} {}
    \NewMathAlphabet{mathbfit} {cmbxti10} {} 
    \NewMathAlphabet{mathbfss} {cmssbx10} {} 
  \fi
  \ifAMStwofonts
    \ifCUPmtlplainloaded \else
      \NewSymbolFont{upmath} {eurm10}
      \NewSymbolFont{AMSa} {msam10}
      \NewMathSymbol{\upi}     {0}{upmath}{19}
      \NewMathSymbol{\umu}     {0}{upmath}{16}
      \NewMathSymbol{\upartial}{0}{upmath}{40}
      \NewMathSymbol{\leqslant}{3}{AMSa}{36}
      \NewMathSymbol{\geqslant}{3}{AMSa}{3E}
           
      \let\oldge=\ge

    \fi
  \fi
\fi 

\ifnfssone
  \newmathalphabet{\mathit}
  \addtoversion{normal}{\mathit}{cmr}{m}{it}
  \addtoversion{bold}{\mathit}{cmr}{bx}{it}
  \newmathalphabet{\mathbfit} 
  \addtoversion{normal}{\mathbfit}{cmr}{bx}{it}
  \addtoversion{bold}{\mathbfit}{cmr}{bx}{it}
  \newmathalphabet{\mathbfss} 
  \addtoversion{normal}{\mathbfss}{cmss}{bx}{n}
  \addtoversion{bold}{\mathbfss}{cmss}{bx}{n}
  \ifAMStwofonts
    \ifCUPmtlplainloaded \else
      \UseAMStwoboldmath
      \makeatletter
      \new@mathgroup\upmath@group
      \define@mathgroup\mv@normal\upmath@group{eur}{m}{n}
      \define@mathgroup\mv@bold\upmath@group{eur}{b}{n}
      \edef\UPM{\hexnumber\upmath@group}
      \new@mathgroup\amsa@group
      \define@mathgroup\mv@normal\amsa@group{msa}{m}{n}
      \define@mathgroup\mv@bold\amsa@group{msa}{m}{n}
      \edef\AMSa{\hexnumber\amsa@group}
      \makeatother
      \mathchardef\upi="0\UPM19
      \mathchardef\umu="0\UPM16
      \mathchardef\upartial="0\UPM40
      \mathchardef\leqslant="3\AMSa36
      \mathchardef\geqslant="3\AMSa3E

      \let\oldge=\ge

    \fi
  \fi
\fi 

\ifnfsstwo
  \DeclareMathAlphabet{\mathbfit}{OT1}{cmr}{bx}{it}
  \SetMathAlphabet\mathbfit{bold}{OT1}{cmr}{bx}{it}
  \DeclareMathAlphabet{\mathbfss}{OT1}{cmss}{bx}{n}
  \SetMathAlphabet\mathbfss{bold}{OT1}{cmss}{bx}{n}
  \ifAMStwofonts
    \ifCUPmtlplainloaded \else
      \DeclareSymbolFont{UPM}{U}{eur}{m}{n}
      \SetSymbolFont{UPM}{bold}{U}{eur}{b}{n}
      \DeclareSymbolFont{AMSa}{U}{msa}{m}{n}
      \DeclareMathSymbol{\upi}{0}{UPM}{"19}
      \DeclareMathSymbol{\umu}{0}{UPM}{"16}
      \DeclareMathSymbol{\upartial}{0}{UPM}{"40}
      \DeclareMathSymbol{\leqslant}{3}{AMSa}{"36}
      \DeclareMathSymbol{\geqslant}{3}{AMSa}{"3E}

      \let\oldge=\ge

    \fi
  \fi
\fi 

\ifCUPmtlplainloaded \else
  \ifAMStwofonts \else 
    \def\upi{\pi}
    \def\umu{\mu}
    \def\upartial{\partial}
  \fi
\fi

\title{The Peculiar Rotation Curve of NGC~157}
\author[S. Ryder, A. Zasov, V. McIntyre, W. Walsh and O. Sil'chenko]
       {S. D. Ryder,$^{1}$\thanks{Present address: Joint Astronomy
        Centre, 660 N. A'Ohoku Place, Hilo, HI 96720, U.S.A. E-mail:
        sryder@jach.hawaii.edu.},
        A. V. Zasov,$^{2}$ V. J. McIntyre,$^{3}$ W. Walsh$^{1,4}$
        and O. K. Sil'chenko$^{2}$\\
        $^{1}$ School of Physics, University of New South Wales, Sydney 2052,
        Australia\\
        $^{2}$ Sternberg Astronomical Institute, Moscow State University,
        Universitetskij Prospect 13, 119899 Moscow, Russia\\
        $^{3}$ School of Physics, University of Sydney, NSW 2006, Australia\\
        $^{4}$ RAIUB, Auf dem Huegel 71, Bonn D-53121, Germany}
\date{Accepted 1997 September 12.
      Received July 7;
      in original form 1997 February 7}

\pagerange{\pageref{firstpage}--26}
\pubyear{1997}

\begin{document}

\maketitle

\label{firstpage}

\begin{abstract}
We present the results of a new \HI, optical, and H$\alpha$
interferometric study of the nearby spiral galaxy NGC~157. Our
combined C- and D-array observations with the VLA show a large-scale,
ring-like structure in the neutral hydrogen underlying the optical
disk, together with an extended, low surface density component going
out to nearly twice the Holmberg radius. Beginning just inside the
edge of the star-forming disk, the line of nodes in the gas disk
commences a 60$^{\circ}$~warp, while at the same time, the rotation
velocity drops by almost half its peak value of 200~\kms, before
leveling off again in the outer parts. While a flat rotation curve in
NGC~157 cannot be ruled out, supportive evidence for an abrupt decline
comes from the ionised gas kinematics, the optical surface photometry,
and the global \HI~profile. A standard `maximum-disk' mass model
predicts comparable amounts of dark and luminous matter within
NGC~157. Alternatively, a model employing a disk truncated at 2~disk
scale lengths could equally well account for the unusual form of the
rotation curve in NGC~157.
\end{abstract}


\section[]{Introduction}

Observations of the neutral hydrogen in `normal' spiral galaxies
have frequently revealed new and surprising aspects of their structure.
Structures in the gas such as warps, tails, and superbubbles point to
a more turbulent history than that indicated by the stellar
distribution. The nearby SAB(rs)bc galaxy NGC~157 is a striking example
of this.  Blue light photographs (Lynds 1974) show a dusty, flocculent
disk of high surface brightness, which is reasonably symmetric and
apparently undisturbed. In contrast, the \HI~aperture synthesis
observations presented here show a kinematic warp and an unusual
rotation curve, which may point to a strongly warped disk or a peculiar
dark matter distribution.

Some hints of NGC~157's distinctive qualities are contained in earlier
optical studies. Blackman (1979) presented photographic {\em UBVR\/}
luminosity profiles, which pointed to the presence of a second, outer
exponential disk, with a longer scale length than the inner disk.  He
also conjectured (from modeling of the spiral arm morphology) that the
rotation curve might be declining at large radii. The disk kinematics
of NGC~157 were first investigated by Burbidge, Burbidge \&
Prendergast (1961), who derived a slowly-rising rotation curve and a
total mass $\sim6\times10^{10}$~M$_{\odot}$ on the basis of two
long-slit spectra close to the galaxy's major axis.  Zasov \& Kyazumov
(1981, hereafter ZK) obtained long-slit spectra of NGC~157 at a number
of position angles, and found good agreement with the Burbidge
et~al. rotation curve in the southwest part of the disk, but a
significant drop in rotational velocity (up to $\sim100$~km~s$^{-1}$)
beyond a radial distance of 55~arcsec to the northeast. Further
evidence of peculiar kinematics comes from the global \HI~profiles of
NGC~157 obtained with the Jodrell Bank 76-m radio telescope by
Staveley-Smith \& Davies (1987; hereafter SSD), and with the Parkes
64-m telescope by Mathewson, Ford \& Buchhorn 1992; hereafter
MFB). Both spectra show a two-horned profile with a strange pair of
shoulder-like features. All this is made the more curious by the fact
that the galaxy is quite isolated. No close companions of comparable
brightness are seen in the POSS images, and the nearest galaxy in the
NED database within $\delta(cz)=\pm500$~\kms~is UGCA~14, at a projected
distance of $3\fdg5$ ($\sim1.3$~Mpc for $D=20.9$~Mpc; Tully 1988). In
fact, there are only five catalogued galaxies within $5^{\circ}$ and
$\delta(cz)=\pm1500$~\kms.

There are also hints from the literature of strong {\em internal\/}
forces acting on the galaxy. At the position of the peculiar motions
seen by ZK, blue light photographs show no obvious signs of a
perturbation or companion object. However, \Halpha\ images (ZK; Rozas,
Beckman \& Knapen 1996) reveal a `ring' of \HII~regions centered
$\sim1$~arcmin northeast of the nucleus of NGC~157 and
some $40''$ ($\sim4$~kpc) in diameter. A ring-shaped feature like this
could be caused by propagating star formation on the periphery of an
expanding \HI~`superbubble'. If so, then (relative to its parent
galaxy) it would be one of the largest and most energetic examples
found to date.

In order to unravel some of these mysteries, and to gain a better
understanding of the morphology and evolution of NGC~157, we have
made the first \HI~map of NGC~157; obtained new wide-field $B$, $I$, and
\Halpha~imaging; and mapped the kinematics in detail of the inner disk
using \Halpha\ Fabry-Perot interferometry. Details of these observations,
and their reductions, are given in Section~\ref{s:obs}. An analysis of the
galaxy's kinematical and morphological parameters is presented in
Section~\ref{s:pars}. Our interpretation of these new results is
discussed in Section~\ref{s:disc}, and a summary of our conclusions
follows in Section~\ref{s:conc}.

\section[]{Observations}\label{s:obs}

\subsection[]{H\,{\sevensize\bf I} Mapping}\label{s:h1m}

NGC~157 was observed with the two lowest-resolution configurations of
the Very Large Array (VLA) of the National Radio Astronomy Observatory
(NRAO).  A total of 3.7~hours on-source integration time was obtained
with the C-array on 1994~October~27, and a further 2.8~hours with the
D-array on 1995~May~26. The primary flux density (and bandpass)
calibrator was 3C48, assumed to have a flux density of 16.12~Jy at
1.413~GHz. The nearby radio source 0022$+$002 (J2000) was used as a
secondary calibrator to monitor variations in the gain and phase. To
be sure of covering the full projected velocity width of NGC~157
(324~km~s$^{-1}$; SSD) at modest resolution (10.4~km~s$^{-1}$ after
on-line Hanning smoothing), the two Intermediate Frequencies (IFs)
were configured to 64~channels each with 1.56~MHz bandwidth. By tuning
the IFs to two closely-spaced frequencies (Table~\ref{t:obspars})
spanning that of the redshifted \HI~line, it was possible to
accommodate the full profile width, with a few channels of continuum
at either end of the band, and with 8~channels of overlap between the
IFs.

\begin{table}
 \caption{VLA observing parameters for NGC~157.}
 \label{t:obspars}
 \begin{tabular}{lc}
Parameter                        & Value \\
\hline
$\alpha$ (J2000 pointing centre) & $00^{\rm h}\, 34^{\rm m}\, 46\fs5$ \\
$\delta$ (J2000 pointing center) & $-08^{\circ}\, 23\arcmin\, 48\arcsec$ \\
Distance adopted                 & 20.9~Mpc \\
Natural-weighted beam FWHP       & $41\arcsec \times 27\arcsec$
                                   ($4.2\times2.8$~kpc) \\
Uniform-weighted beam FWHP       & $18\arcsec \times 12\arcsec$
                                   ($1.8\times1.2$~kpc) \\
Channel map rms noise (natural)  & 0.6~mJy~beam$^{-1}$ \\
Channel map rms noise (uniform)  & 1.3~mJy~beam$^{-1}$ \\
Continuum map rms noise (uniform) & 0.3~mJy~beam$^{-1}$ \\
IF~1 central frequency             & 1412.9~MHz \\
IF~2 central frequency             & 1411.6~MHz \\
Channel increment (unsmoothed)    & 5.2~\kms \\
 \end{tabular}
\end{table}

The C-array and D-array $uv$-datasets were processed separately using
implementations of the NRAO {\sc aips} software at the NRAO Array
Operations Centre and at the Australia Telescope National
Facility. The visibility data were calibrated using standard VLA
procedures, and bad data removed interactively. After combining the
two datasets in the $uv$-plane, the imaging task MX was used to construct
a cube of dirty channel maps for each of the two IFs, as well as cubes
of the corresponding dirty beam shapes. At this stage, the two cubes were
`sliced' into their constituent channel maps, and then a `supercube'
was constructed by stacking all available maps in order of increasing
frequency. After dropping some channel maps at either end of the IF
bandpass with low signal-to-noise ratio, we were left with 112 usable
channel maps in the supercube spanning 581~\kms.

A total of 20~channels to either side of the main profile containing
only continuum emission were averaged together to form a master dirty
continuum image, and then this image was subtracted from every map in
the supercube.  Although continuum subtraction in the $uv$-plane is
always preferable over subtraction in the image domain (Killeen 1993),
the need to combine two IFs made this impractical. In any case, since
$\Delta\nu/\nu_{0}\sim0.002$, and the continuum emission from NGC~157
itself is not particularly strong, this method of continuum
subtraction worked extremely well.

Because they highlight different structural features of NGC~157, two
distinct supercubes were constructed using different weighting schemes
in the imaging process. Uniform weighting yields the best possible
spatial resolution, while natural weighting offers improved
sensitivity to extended, low surface brightness emission, both of
which turn out to be important in understanding the nature of
NGC~157. In each case, the pixel scale was chosen to allow at least 3
pixels across the synthesized beam minor axis (Table~\ref{t:obspars}),
and the map size to match the VLA primary beam at 1.4~GHz
($\sim30$~arcmin). By comparing the total CLEANed flux density with
the improvement in rms noise obtained, as a function of the number of
CLEAN iterations performed using the APCLN task, we elected to perform
200 CLEAN iterations on each map of the natural-weighted supercube,
while the uniform-weighted data was found to require no further
CLEANing.

Following primary-beam corrections, and conversion from frequency to
optical heliocentric velocity, moment maps (\HI~column density,
velocity, and velocity width) were constructed from each of the
two supercubes, using masks derived from a spatially and spectrally
smoothed version of the supercube, and by setting a flux cutoff that
just enabled a clean separation of the \HI~emission from background
noise. Beam full-widths at half power (FWHP) and rms noise levels in
the final CLEANed maps are summarised in Table~\ref{t:obspars}.

\subsection[]{Optical Imaging}

NGC~157 was observed under photometric conditions with the Kitt Peak
National Observatory 0.9-m telescope on 1995~July~3~UT. A Tektronix
$2{\rm K}\times2{\rm K}$~CCD was used at $f/7.5$ with the doublet
corrector, making available a field of 23.2~arcmin $\times$ 23.2~arcmin at
0.68~arcsec~pixel$^{-1}$. Two 500~s exposures in $B$, and two 300~s
exposures in $I$ were obtained, together with short bracketing
exposures of Landolt's (1983) Field~109. On the previous night,
NGC~157 was observed once for 1000~s in the KP1564 (6618~\AA\ central
wavelength, 74~\AA\ FWHM) \Halpha\ filter, and also for 1000~s in the
KP808 (6411/88) red continuum filter. Images of Cygnus OB2 No.~9
(Massey et al. 1988) with the KP1564 filter served as a flux
calibration for the \Halpha\ data.

Following removal of the bias level and structure, attempts were made
to flatfield the data using dome flatfields, twilight sky flatfields,
or both, but residual structure at the 5 per cent level was still
apparent near the edges of the frames. Finally, low-order polynomial
fits to these border regions were computed and used to normalise both
the object and calibration frames. Aperture photometry of the
broadband standards over a large range in airmass allowed us to
compute photometric zero points, extinction, and colour-correction
terms. Each pair of $B$ and $I$ images of NGC~157 were averaged
together, after scaling to a common modal value in the central
regions, and registration using field stars. The red continuum image
of NGC~157 was convolved to the same stellar FWHM as the \Halpha\
image, registered using field stars, multiplied by a factor of 1.2
(the ratio of the continuum to line filter full widths), and then
subtracted from the KP1564 image, leaving just the pure \Halpha\
emission.

\subsection[]{H$\alpha$ Fabry-Perot Interferometry}

The two-dimensional velocity field for the ionized gas in NGC~157 was
obtained at the 6-m telescope at the Special Astrophysical Observatory
of the Russian Academy of Sciences on 1995~October~24, when the seeing
averaged 2.5~arcsec. A scanning Fabry-Perot interferometer was
installed in the pupil plane of a focal reducer which was attached to
the $f/4$ prime focus of the telescope (yielding an $f/2.4$ output
beam). The detector used was a $512\times512$ pixel intensified
photon-counting system. The interferometer was operated in the 501st
order at 6562.8~\AA, and a narrow-band (10~\AA) filter at the
appropriate H$\alpha$ redshift served to block adjacent orders.  A
total of 32 images (each 4~arcmin $\times$ 4~arcmin) were obtained with a
velocity increment of 18.84~km~s$^{-1}$, with 3~minutes of integration
at each velocity step. A neon emission line at 6598.95~\AA\ was used
for wavelength calibration. The raw observational data were
rebinned to cubes of $32 \times 256 \times 256$ pixels, with an image
scale of 0.92~arcsec~pixel$^{-1}$ and a spectral resolution of $2-2.5$
channels ($40 - 50$~km~s$^{-1}$). Following the steps outlined
in Le~Coarer et al. (1992) and Laval et al. (1987), standard reduction
procedures (corrections for phase shifting, subtraction of the night-sky
emission-line spectrum, construction of the velocity field, etc.) were
performed using the {\sc ADHOC} Fabry-Perot software package developed
at the Marseille Observatory (Boulesteix 1993).

\section[]{Results and Analysis}\label{s:pars}

\subsection[]{H\,{\sevensize\bf I} distribution}
\label{s:h1dist}

\begin{figure*}
\vspace{22cm}
\includegraphics{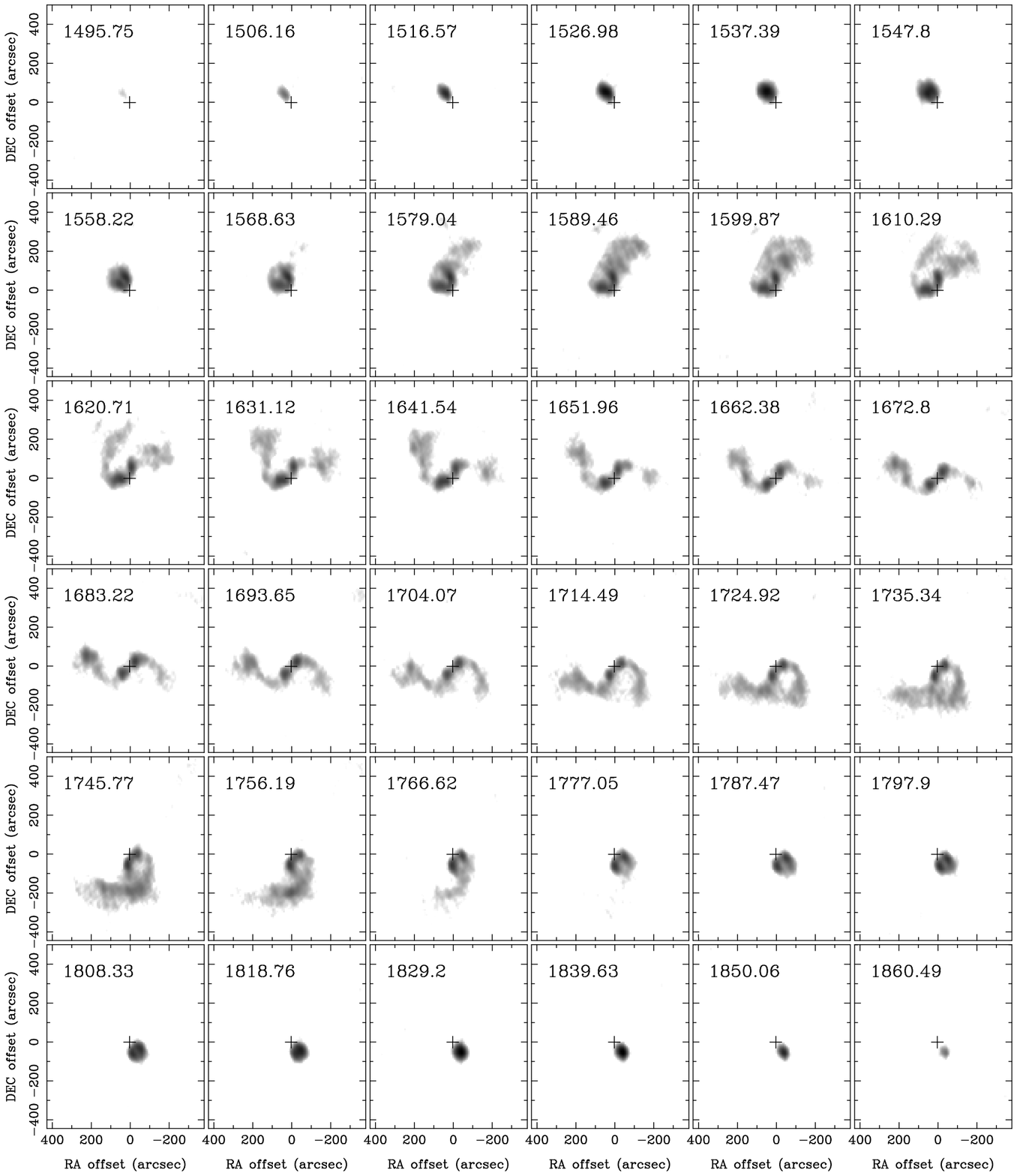}
\caption{\HI\ channel maps of NGC~157 at
$\sim10$~km~s$^{-1}$ intervals from our natural-weighted `supercube'.
Heliocentric velocities in the optical convention are given for each
map in units of km~s$^{-1}$. The flux range plotted is from 2 to
40~mJy~beam$^{-1}$, and is displayed logarithmically to enhance
details in the extended \HI\ disk. The cross in each panel indicates
the position of the dynamical center (coincident with the optical
nucleus).}  \label{f:n157cm}
\end{figure*}

The resulting channel maps from the natural-weighted supercube are
shown in Figure~\ref{f:n157cm}. Already, there are indications from
this sequence of maps of some peculiarities in the \HI\ structure and
dynamics; namely the existence of an extended gas disk beyond the
optical radius ($r_{25}=2.08$~arcmin; de~Vaucouleurs et~al. 1991,
hereafter RC3) of this galaxy, and from the bending of the isovelocity
features, an associated warping of the gas disk.

\begin{figure*}
\vspace{22cm}
\includegraphics{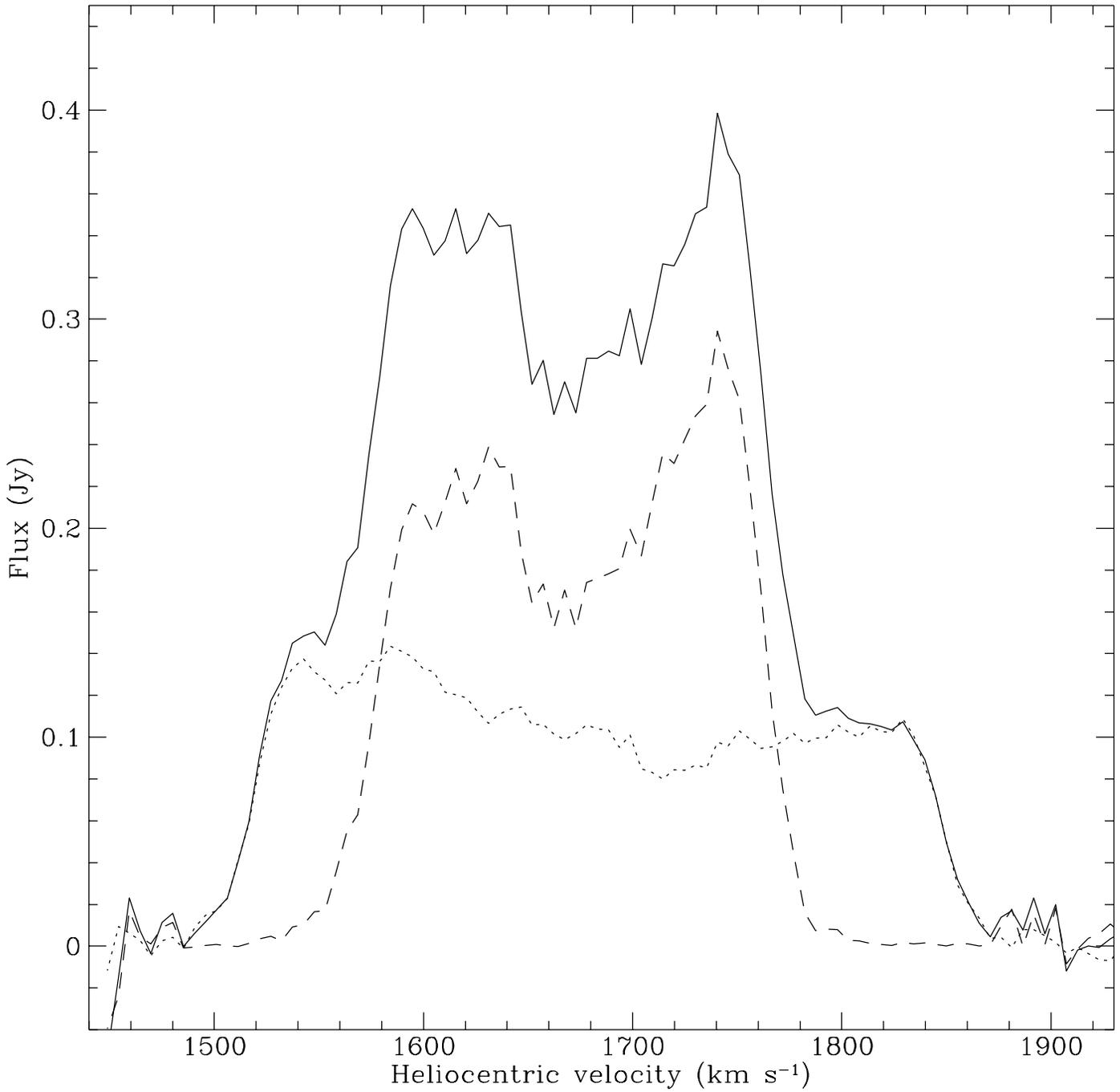}
 \caption{Global \HI\ profile (solid line) for NGC~157, obtained by summing
 the emission within each channel map of the natural-weighted `supercube'.
 The dotted profile is the emission from just the inner disk, while the dashed
 profile represents the emission from the outer, warped disk.}
 \label{f:h1prof}
\end{figure*}

Before considering this disk in more detail, it is prudent to examine
the extent to which we may be losing flux and large-scale structure
due to missing baselines. To do that, we have computed the global \HI\
profile for NGC~157 by using a moving window to optimise the emission
measured within each channel map of the natural-weighted supercube,
and plotted the fluxes versus the respective map velocity in
Figure~\ref{f:h1prof}.  Our profile matches well the general form of
those published by SSD and by MFB, particularly the `shoulders' at
$\sim1540$~km~s$^{-1}$ and $\sim1800$~km~s$^{-1}$, although our peak
measured fluxes are slightly higher. A comparison of our profile
moments (integrated flux, $W_{20}$, etc.) with those from SSD and MFB
appears in Table~\ref{t:profpars}. We take the good agreement shown
there as evidence that our VLA observations are not resolving out a
significant low column density component, at least within the
$12-15$~arcmin single dish beamwidths.

The shape of the spectrum can be shown to be the result of the
superposition of two spatially distinct components in the gas
distribution. The dotted and dashed lines in Figure~\ref{f:h1prof} are
spectra that were obtained by separately summing the regions inside
and outside, respectively, of the $\mu_{B}=25$~mag~arcsec$^{-2}$
isophote. A similar result is obtained if the summation is done using
a mask chosen to separate the `warped' component of the \HI\
(Section~\ref{s:kin}). The flux ratio of the two components is ${\rm
(inner/outer)}=(37\pm2~{\rm Jy}~{\rm km}~{\rm s}^{-1})/ (41\pm2~{\rm
Jy}~{\rm km}~{\rm s}^{-1})=0.9$. Their systemic velocities differ
slightly, being $1674\pm3$ and $1685\pm3$~\kms\ respectively.

A flux integral of $(78\pm4)$~Jy~\kms\ corresponds to a
total mass of \HI\ of $(7.9\pm0.4)\times10^{9}$~M$_{\odot}$ for the
distance of 20.9~Mpc (Tully 1988) adopted here. Taking the absolute
blue magnitude for NGC~157 to be $M_{B}=-20.8$ (Section~\ref{s:sphot}),
we therefore find $M_{\rm HI}/L_{B}\sim0.23$, quite typical of an
Sbc-type galaxy (Roberts \& Haynes 1994).

\begin{table*}
\begin{minipage}{120mm}
 \caption{Global \HI\ profile parameters for NGC~157.}
 \label{t:profpars}
 \begin{tabular}{lr@{$\pm$}lr@{$\pm$}lr@{$\pm$}l}
Parameter                     & \multicolumn{2}{c}{Staveley-Smith \&
Davies 1987} & \multicolumn{2}{c}{Mathewson et al. 1992} &
\multicolumn{2}{c}{This Study} \\
\hline
Flux Integral (Jy~km~s$^{-1}$) & 62.6~ & ~5.5 & 72.8~ & ~7.1 & 78.0~ & ~4.0 \\
${V_{\rm hel}}^{a}$ (km~s$^{-1}$)   & 1669~ & ~7 & 1655~ & ~7 & 1682~ & ~12 \\
$W_{20}$ (km~s$^{-1}$)        & 324~  & ~14  & 336~ & ~14 & 326~  & ~10 \\
\end{tabular}
\medskip
~~\\
$^{a}$Mean of velocities where the \HI\ profile falls to 20~per cent of the
peak flux (Staveley-Smith \& Davies 1987; this study), or 50~per cent of the
peak flux (Mathewson et~al. 1992)
\end{minipage}
\end{table*}

\begin{figure*}
\vspace{22cm}
\includegraphics{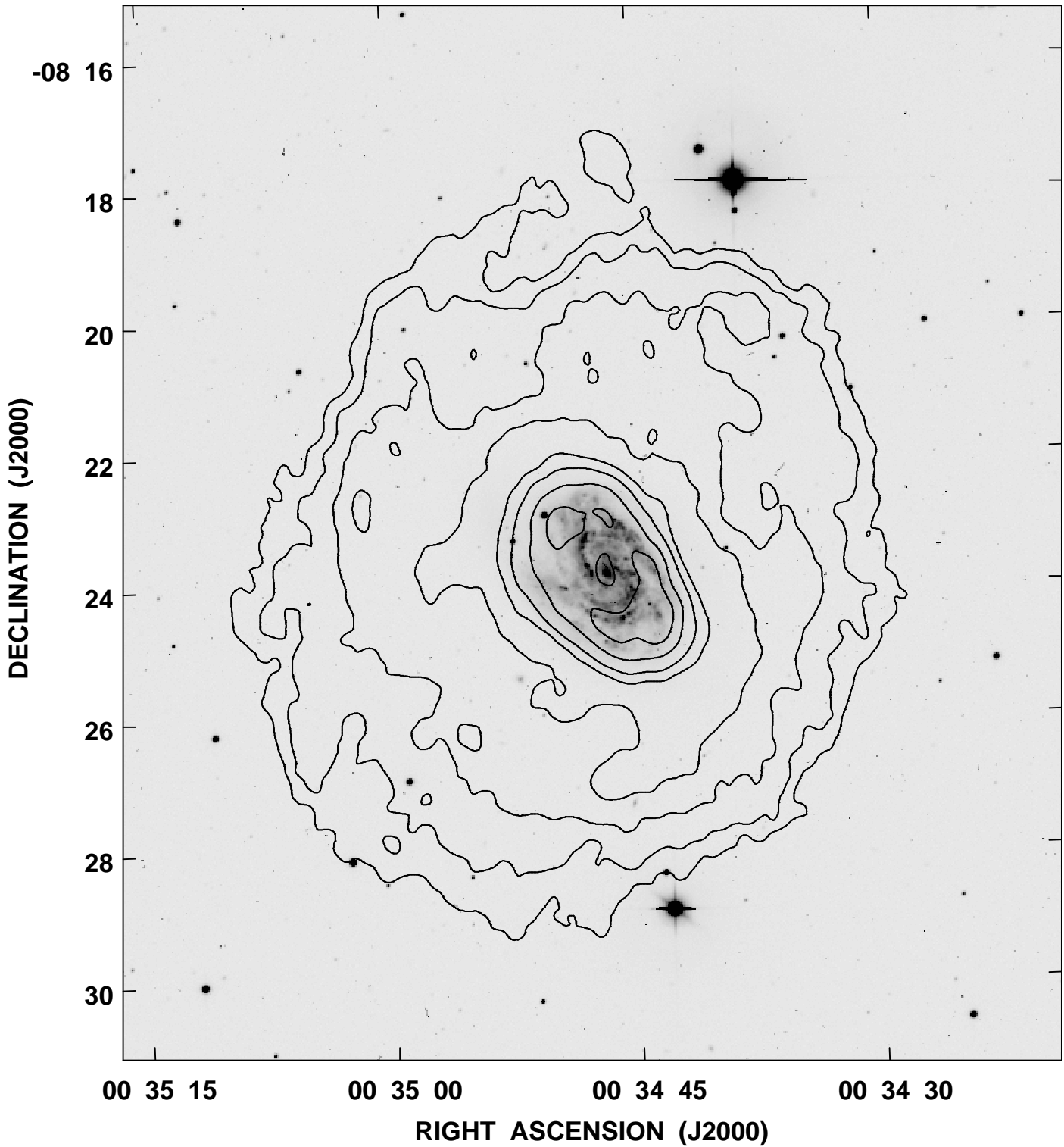}
 \caption{Contours of \HI\ column density overlaid on a $B$-band image from
the KPNO 0.9-m telescope. The contours correspond to (projected) column
densities of 0.5, 1.0, 2.0, 3.5, 7.0, 10.0, 15.0, and $20.0\times10^{20}
$~cm$^{-2}$.}
 \label{f:nah1onb}
\end{figure*}

\begin{figure*}
\vspace{22cm}
\includegraphics{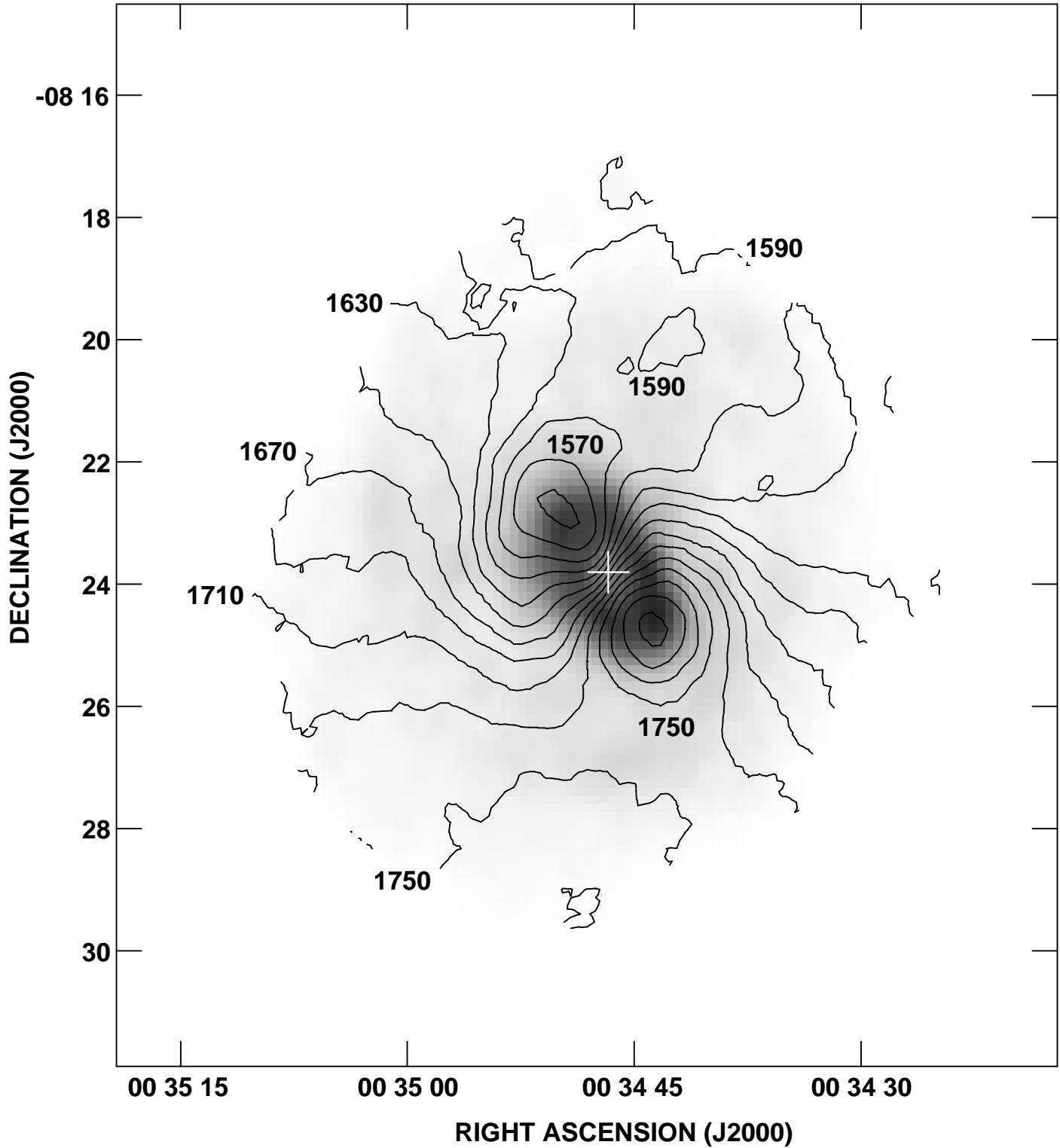}
 \caption{Isovelocity contours plotted on a grey-scale of the same \HI\
distribution as shown in Figure~\protect{\ref{f:nah1onb}}. The contour
interval is 20~km~s$^{-1}$, and major contours are marked. The cross marks
the position of the dynamical center, as determined from the kinematical
analysis.}
 \label{f:nah1vf}
\end{figure*}

The \HI\ column density distribution (zeroth moment) is shown as
contours superimposed upon a section of the KPNO 0.9-m $B$-band image
in Figure~\ref{f:nah1onb}, and as a grey-scale in
Figure~\ref{f:nah1vf}.  The impression of an extended \HI\ disk in
NGC~157 gained from the channel maps is confirmed by these images,
with the outermost contour, representing a projected \HI\ column
density of $0.5\times10^{20}$~cm$^{-2}$ (0.4~M$_{\odot}$~pc$^{-2}$)
spanning some 11~arcmin, or over 2.5 times the optical diameter. The
flux cutoff imposed in defining the edge of the \HI\ disk is roughly
at $3\times$ the noise level of the moment fitting; however, in
blanking off the low signal-to-noise regions, we were guided also by
the smooth continuity of the velocity field in the outer disk
(Figure~\ref{f:nah1vf}).  NGC~157 possesses a central \HI~`hole' of
the sort commonly seen in spiral galaxies, and the projected gas
column density reaches a peak in a ring underlying the spiral arms. An
arm or tail-like feature appears to be breaking free from the northern
edge of the \HI~disk in Figure~\ref{f:nah1vf}.


The apparent offset between the major axis position angles of the
optical and gas disks in Figure~\ref{f:nah1onb} seems to be the result
of a change in the orientation of the disk orbits between the inner
and outer disk. The isovelocity contours in Figure~\ref{f:nah1vf} show
the kinematic line-of-nodes precesses through $\sim60^{\circ}$ between
the edge of the optical disk and the limits of the outer gas disk.  At
this resolution, the inner disk displays the characteristic pattern
for an inclined disk in differential rotation, albeit with an abrupt
turnover in the projected velocity as a function of radius (discussed
in Section~\ref{s:tover}). In the outer disk, the isovelocity contours
meander in a fashion typical of galaxies with a warp. There are also
kinks in these contours consistent with the action of a density wave
as mentioned earlier.

\begin{figure*}
\vspace{22cm}
\includegraphics{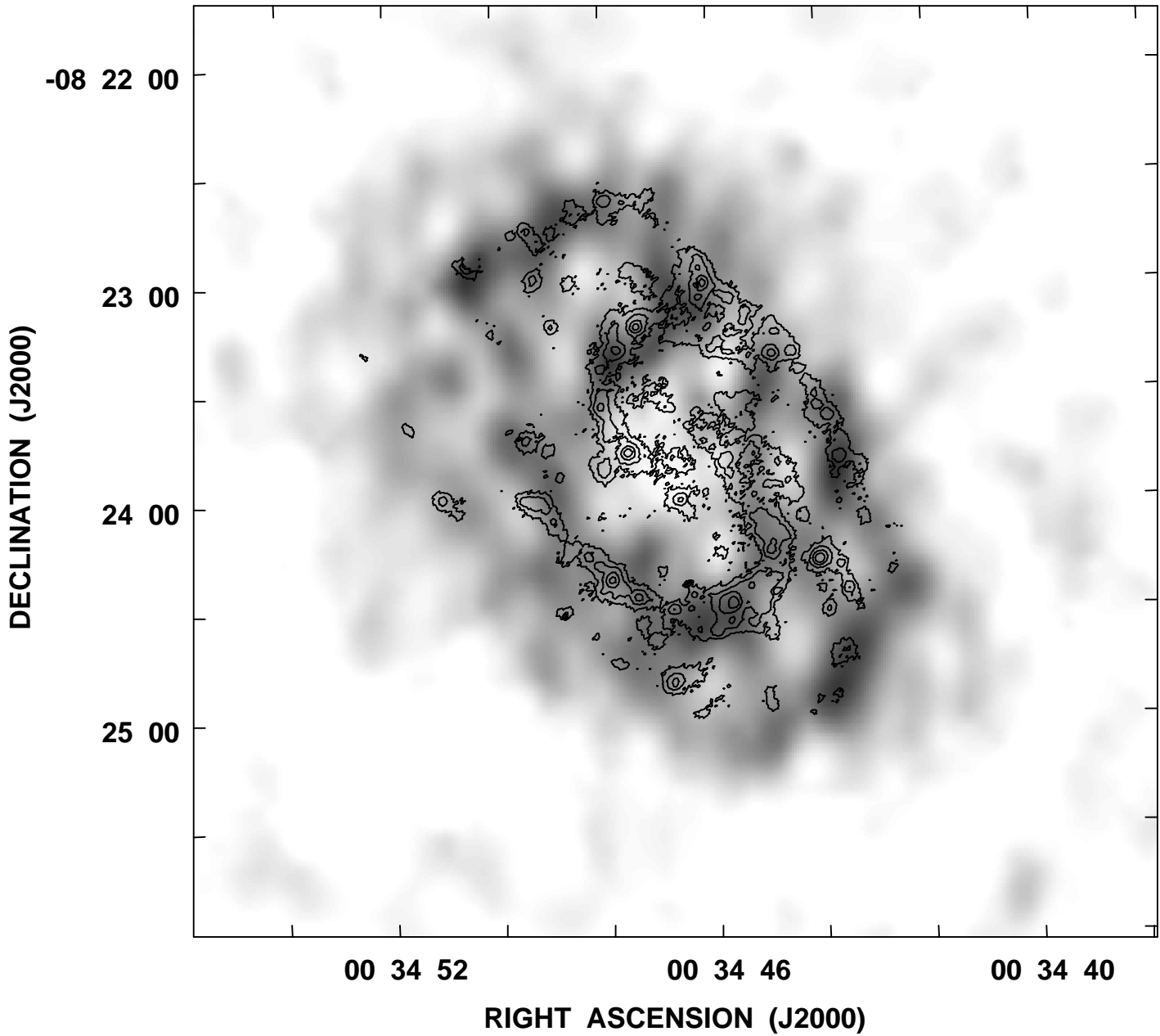}
 \caption{Higher resolution grey-scale image of the uniform-weighted \HI\
in the inner disk of NGC~157, on which contours of the \Halpha\ emission
has been overlaid. Notice the close, but not perfect, coincidence of star
formation activity with peak gas surface density.}
 \label{f:haonh1}
\end{figure*}

A higher resolution map made with uniform-weighting
(Figure~\ref{f:haonh1}) resolves out nearly all of the extended outer
\HI\ disk, but does highlight some interesting structure within the
inner disk. The inner gas `ring' is now revealed to be composed of a
honeycomb-like lattice of rings and cells. Although initially chaotic
in appearance, the peaks in the gas surface density making up the
ridge-lines of these `cells' are shown to be intimately associated
with the spiral arms and other sites of massive star formation, when
contours of the \Halpha\ surface brightness are superimposed. The
\Halpha\ ring to the northeast of the galaxy discussed by ZK is seen
to have a distinct \HI\ counterpart, as do many of the other major
star-forming complexes in NGC~157. At the highest resolution offered
by the \HI\ data, the individual cells have diameters between
2 and 4~kpc, and deprojected \HI\ column densities peaking at
$\sim13$~M$_{\odot}$~pc$^{-2}$ on the cell `walls', dropping to as low
as $1.5$~M$_{\odot}$~pc$^{-2}$ at the centres.  The correlation of
star formation activity with local gas surface density in NGC~157 is
somewhat better than that typically observed in spiral galaxies (e.g.,
Ryder et~al. 1995; Rownd et~al. 1994), and is more like that observed
in the shells of the dwarf irregular galaxy Holmberg~II (Puche
et~al. 1992), although there is no suggestion that the gas `cells' in
NGC~157 are expanding coherently.

\begin{figure*}
\vspace{22cm}
\includegraphics{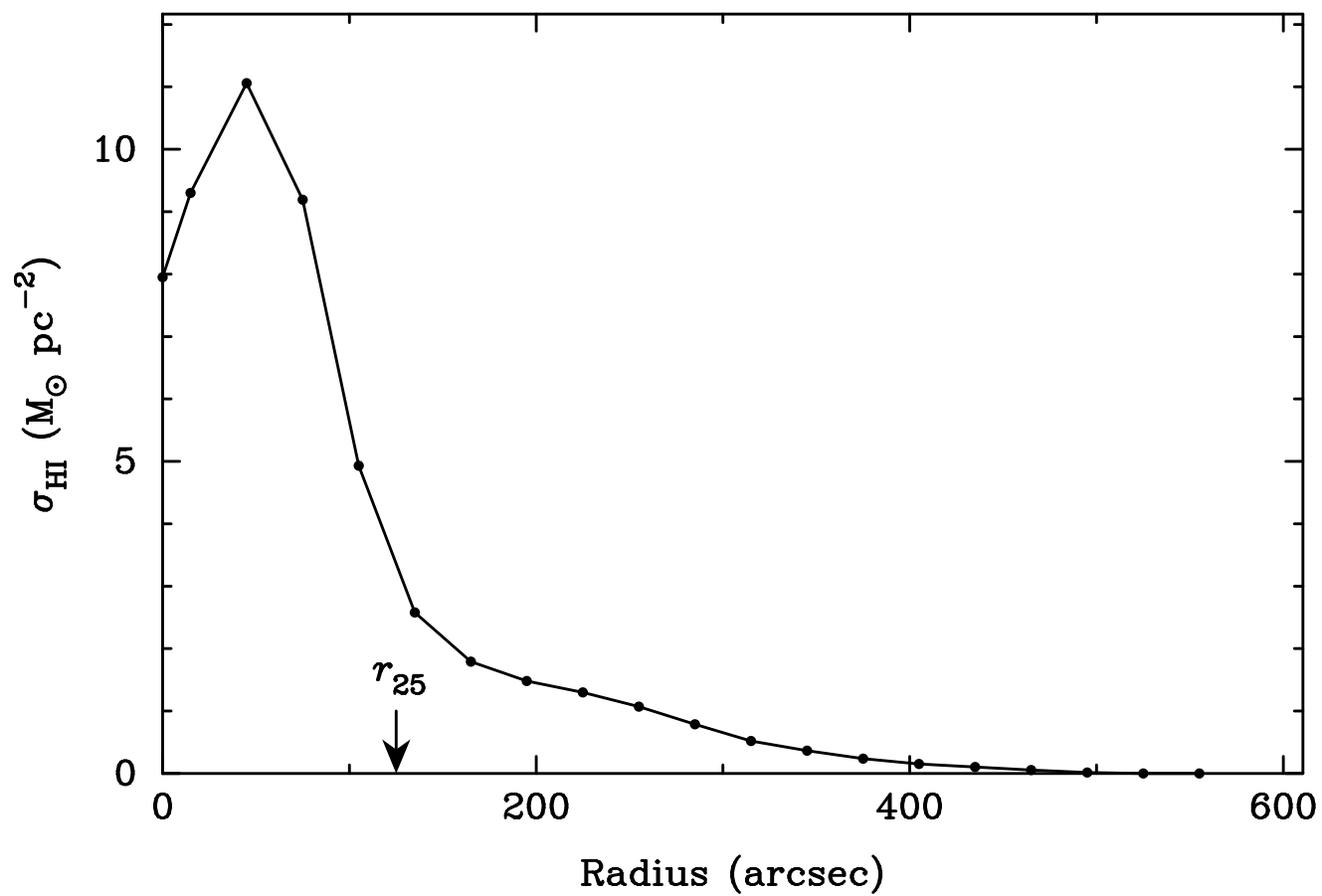}
 \caption{Radial variation of the depojected surface density of \HI\ in
NGC~157, assuming a fixed inclination $i=45^{\circ}$ and position angle
$\theta=220^{\circ}$. The 25~$B$~mag~arcsec$^{-2}$ radius is marked by
the arrow.}
 \label{f:iring}
\end{figure*}

The azimuthally-averaged column density of \HI\ as a function of
radius in the naturally-weighted zeroth moment map is plotted in
Figure~\ref{f:iring}. Since the intention here is to indicate the column
density of gas in the disk of the galaxy that would be seen from
`face-on', we have adopted a fixed inclination ($i=45^{\circ}$ based
on the results of Sections~\ref{s:kin} and \ref{s:sphot} as well as
Grosb\o l (1985)) and position angle ($\theta=220^{\circ}$) and then
calculated the deprojected column densities accordingly.
The arrow in Figure~\ref{f:iring} indicates the radius corresponding
to the $\mu_{B}=25$~mag~arcsec$^{-2}$ isophote ($r_{25}$),
and just on half of the total \HI\ content of NGC~157 lies {\em
outside\/} this radius. The point at which the \HI\ surface density
falls below 1.0~M$_{\odot}$~pc$^{-2}$ equates to $2.1 r_{25}$, an
\HI-to-optical extent bettered by only 5 of the 23 comparable galaxies
studied by Broeils \& van~Woerden (1994; hereafter BvW).

\subsection[]{H\,{\sevensize\bf I} kinematics}\label{s:kin}

We determined a rotation curve from the velocity field by using an
{\sc aips} implementation of the ROTCUR algorithm (Begeman 1989) to
fit rings of variable inclination and position angle under
the assumption of uniform circular motion in each ring.  To begin
with, we analysed only the regions within the optical extent of
NGC~157 using our highest resolution (uniform-weighted) velocity
field, sampled at approximately one synthesized beamwidth (15~arcsec)
intervals. By adopting starting values for the kinematical parameters
(dynamical center position, systemic velocity, and disk tilt and
position angle) drawn from RC3 and our own surface photometry
(Section~\ref{s:sphot}), and allowing all of these parameters to vary
freely, we arrived at the following values for the dynamical center:
$V_{\rm hel}=1676\pm9$~km~s$^{-1}$, $\alpha_{\rm dyn}({\rm J2000}) =
00^{\rm h}\, 34^{\rm m}\, 46\fs7\, (\pm4\arcsec)$, $\delta_{\rm
dyn}({\rm J2000}) = -08^{\circ} \, 23\arcmin\, 49\arcsec\,
(\pm3\arcsec)$. These results are in close agreement with the position
of the optical centre (shown as the pointing centre in
Table~\ref{t:obspars}) and the systemic velocities from the midpoint
of the global \HI\ profile (Table~\ref{t:profpars}).

The analysis was then repeated with the dynamical center held fixed at
these values, so that the combinations of rotational velocity,
inclination, and position angle (kinematic line-of-nodes) which
minimised the systematic velocity residuals within each ring could be
studied. Fixing the dynamical centre also allows us to compare the
results from fitting the entire disk with those that come from fitting
only the `receding' side (i.e., mostly the southern half) or the
`approaching' (northern) side, which then highlights any asymmetries
in the disk kinematics. This same procedure can be applied to the full
\HI\ disk kinematics as revealed by the natural-weighted data,
although at a somewhat coarser resolution (35~arcsec). The dynamical
centre in this case is found to lie within 2~arcsec and 1~\kms\ of
that found earlier, so has been fixed at identical values.

The end result of both of these analyses is shown in
Figure~\ref{f:rotfig}, in which points within 120~arcsec of the centre
are drawn from the uniform-weighted fit, and those further out are
from the natural-weighted data. Clearly, the continuity between these
two is extremely good, and the major differences are in fact between
the two halves of the disk. The principal feature of the rotation
curve is a strong decline that begins just inside $r_{25}$. The exact
shape of the decline is somewhat uncertain because the fits to the
northern (receding) half failed to converge in the transition region,
and the two halves of the disk give rather different results for
$r=200-300$~arcsec. However, the fits to the full disk and southern
half do agree in the transition region, and all three fits agree for
$r>350$~arcsec. At this point, the circular velocity seems to have
settled to a constant value at $\sim120$~\kms\ (55\% of the maximum
rotational velocity), and the inclination is only slightly higher
than that attained in the inner disk.

Although the steady progression in position angle is characteristic of a
warp, the rotational velocity drops by almost half in less
than 1~arcmin, while the inclination varies by no more than
$10^{\circ}$ in this region. The drop in velocity is even faster than
Keplerian for $120<r<190$~arcsec, though such behaviour is permissible
for a disk with a sharp edge (Section~\ref{s:trunc}).

\subsection[]{Ionised Gas Kinematics}\label{s:igk}

\begin{figure*}
\vspace{22cm}
\includegraphics{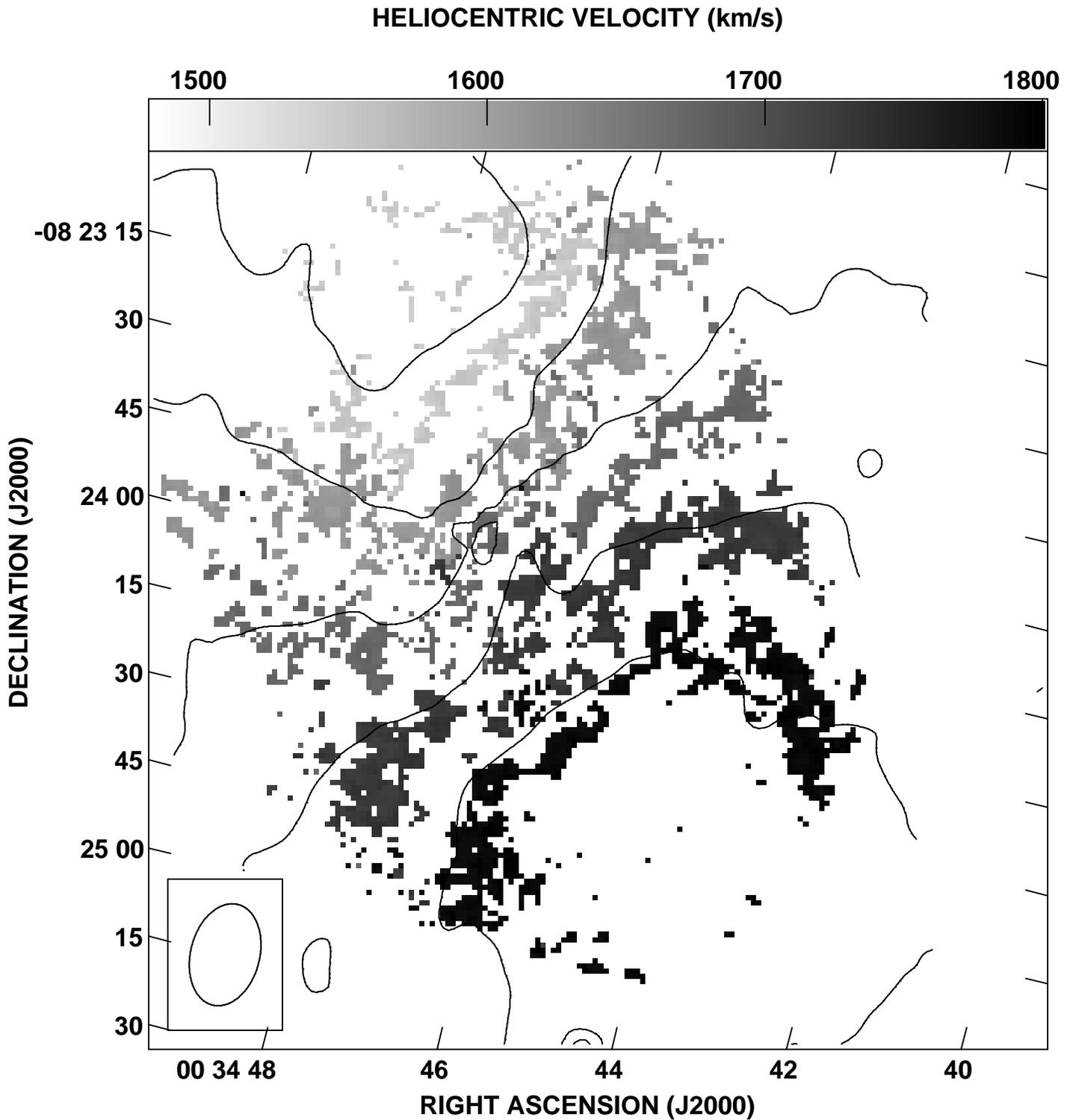}
 \caption{Comparison of the velocity fields obtained from \Halpha\ Fabry-Perot
interferometry (grey-scale) and \HI\ aperture synthesis (contours). For
clarity, only the \Halpha\ velocity ranges $1550\pm10$, $1610\pm10$,
$1670\pm10$, $1730\pm10$, and $1790\pm10$~\kms\ are shown. The contours
correspond to \HI\ velocities of 1550 (northeast), 1610, 1670, 1730,
and 1790~\kms\ (southwest). The synthesised beamsize of the \HI\ data is
shown by the ellipse in the lower-left corner; the spatial resolution of the
\Halpha\ data is $\sim5$ times better.}
 \label{f:vfcomp}
\end{figure*}

Figure~\ref{f:vfcomp} shows a comparison of the \Halpha\ Fabry-Perot
velocity field with the highest resolution velocity field available
from the uniform-weighted \HI\ data. Owing to the differing resolutions,
and the complexity of the \Halpha\ velocity field, only selected velocity
ranges and isovelocity contours for the \HI\ data are shown. There appears
to be a discrepancy between the velocity scales of perhaps $10-20$~\kms,
but this is within the spectral and spatial resolutions of the two
datasets. Otherwise, the agreement between the neutral and ionised
gas kinematics in tracing structure within the inner velocity field
is remarkably good.

A rotation curve was derived from the two-dimensional velocity field
using the ROTCUR algorithm described in Section~\ref{s:kin}.
Preliminary fits indicated a rotation centre again coincident with the
optical nucleus, but a systemic velocity of $1671\pm7$~\kms,
consistent with the velocity offset just mentioned and still within
the range of \HI\ systemic velocities in Table~\ref{t:profpars}.
Keeping these parameters fixed, the best-fitting combination of
rotation velocity, inclination and position angle have been evaluated,
and plotted as solid lines in Figure~\ref{f:rotfig}. Unusual structure
in the nuclear velocity field, possibly associated with a bar
(Section~\ref{s:bar}), makes fitting the rotation curve there difficult,
but in the range $30 < r < 70$~arcsec, consistency between the \Halpha,
\HI, and photometric orientation parameters is excellent.


The \Halpha\ rotation curve rises relatively slowly, as a higher
inclination is initially favoured, but then accelerates as the
inclination drops toward the $\sim50^{\circ}$ angle of the inner disk.
It then levels off at the same peak velocity ($\sim200$~\kms) attained
by the \HI. A bulge component may be responsible for the steep
central velocity gradient. It is worth noting, however, that the inner
maximum of the rotation curve cannot be accounted for by a combination
of a King-profile bulge (even if it is sharply truncated) and the
disk, but can be fit by the combination of a steep, inner exponential
disk and a flatter, outer disk (Blackman 1979; Section~\ref{s:trunc}).

%

\subsection[]{1.4~GHz continuum}

\begin{figure*}
\vspace{22cm}
\includegraphics{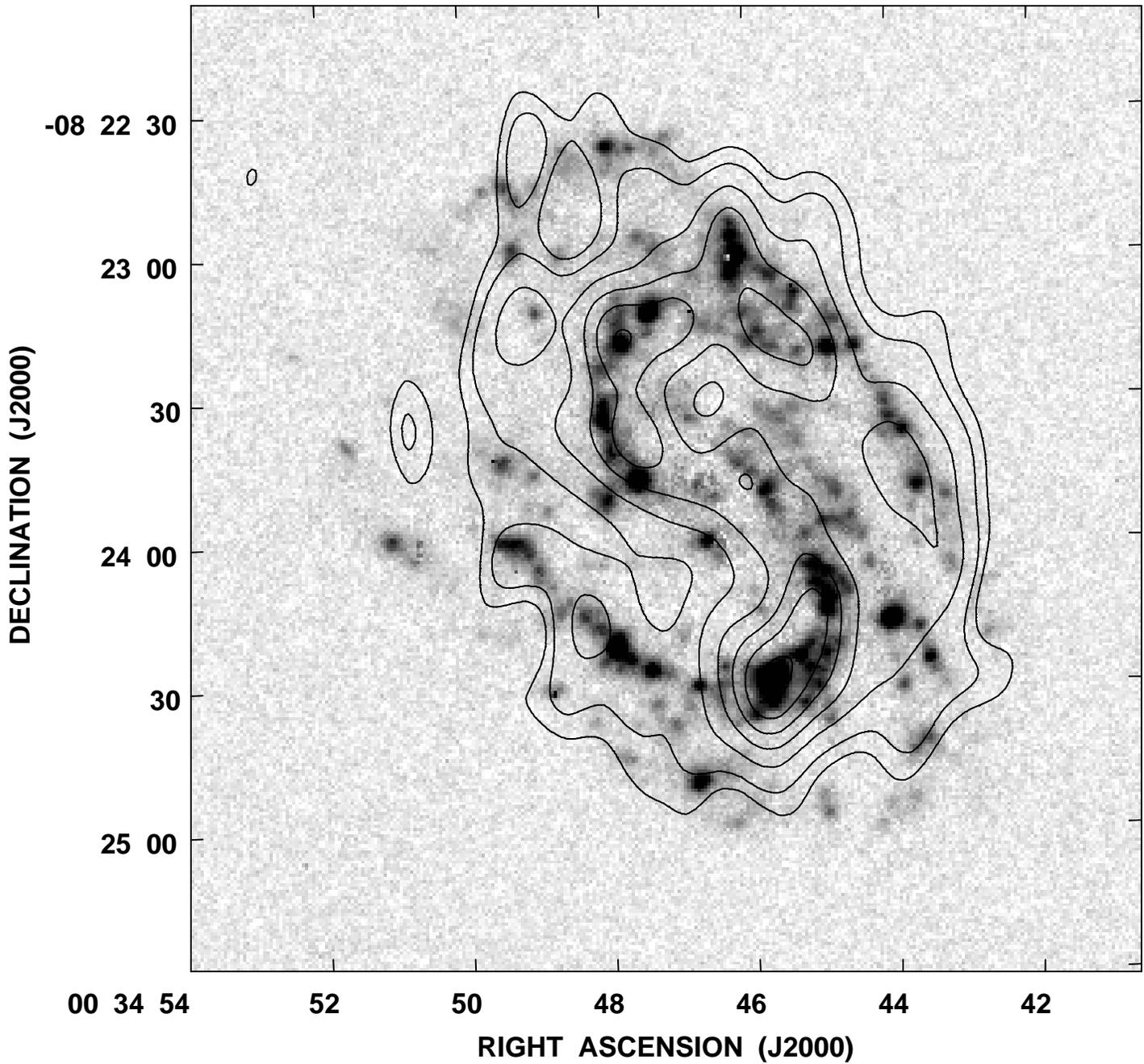}
 \caption{Contours of the uniform-weighted 1.4~GHz continuum overlaid on
the pure \Halpha\ image of NGC~157. The \Halpha\ surface brightness is
displayed logarithmically to bring out the diffuse component, while the
contours correspond to flux densities of 1.5, 2.0, 3.0, 4.0, 5.0, 6.0,
and 7.0~mJy~beam$^{-1}$.}
 \label{f:contonha}
\end{figure*}

Since only one of the IFs contained a significant number of line-free
channels, the {\sc aips} task MX was used for both imaging and
CLEANing of the 1.4~GHz continuum (in preference to the continuum map
yielded as a by-product in Section~\ref{s:h1m}, which was formed by
averaging in the image plane of non-contiguous channels, and which
could only be CLEANed over a limited area). A total of 17 channels was
extracted from IF~2 of the $uv$-dataset, Fourier transformed to the
image domain using uniform-weighting, and then subjected to 1000
iterations of the CLEAN algorithm using a similarly-extracted dirty
beam.  Contours of the 1.4~GHz continuum emission have been
superimposed on top of the \Halpha\ image of NGC~157 in
Figure~\ref{f:contonha}.

As was the case with the \HI\ (Section~\ref{s:h1dist}), there is a
close, though not strict correspondence of the radio continuum flux
density with the \Halpha\ surface brightness. A continuous ridge of
emission underlies both the major spiral arms, passing through the
nucleus and reaching a peak midway along the southern arm, but with an
abrupt break shortly thereafter. Diffuse emission is seen to fill the
entire optical disk, but does not extend much beyond this. All of this
is consistent with the origin of the continuum emission being mostly
from relativistic electrons diffusing along galactic magnetic field
lines, but with some contribution from the thermal component from
\HII~regions. Interestingly, the radio continuum contours conform also
to the shape of the northern ring seen in \HI\ and in \Halpha.

An unresolved radio continuum source of some 146~mJy lies just over
5~arcmin southeast of the nucleus of NGC~157. It has no counterpart in
existing radio or optical catalogues. The source lies well outside the
optical disk, but is just inside the detectable limit of the \HI\
disk.
Because of the high galactic latitude of NGC~157 ($b=-71^{\circ}$),
this object is almost certainly at a greater redshift than
NGC~157. This opens up the tantalising possibility of measuring \HI\
absorption, and thus the gas spin temperature, in a part of the gas
disk where such measurements are rare.

We have searched for signs of such absorption in both the full
$uv$-dataset, and in the continuum-subtracted cube, but to no
avail. Such a null result is perhaps to be expected, given the
conclusion of Dickey, Brinks \& Puche (1992) that even the longest
baselines of the VLA C- and D-arrays (3~km maximum) are inadequate for
resolving out the \HI~emission fluctuations across the continuum
source. Followup observations with the VLA A- or B-array may yet
however succeed in placing limits on the gas temperature in the very
outer parts of the gas disk in NGC~157.

\subsection[]{Surface photometry}\label{s:sphot}

The {\sc gasp} software package was used to compute radial surface
brightness profiles and disk orientation parameters for NGC~157 in a
manner similar to that described by Ryder \& Dopita (1994) for a large
sample of southern spiral galaxies. The $B$, $I$, and
continuum-subtracted \Halpha\ images were first block-averaged to
1.36~arcsec~pixel$^{-1}$, after which the cores and extensive halos of
the two bright foreground stars visible in Figure~\ref{f:nah1onb} were
masked out. The sky background was determined from the modal peak of
the histogram of all data values within 20 pixels of the image border,
and found to be $22.01\pm0.02$ and $19.28\pm0.01$~mag~arcsec$^{-2}$ in
$B$ and $I$ respectively.

For a given semi-major axis, {\sc gasp} iteratively attempts to fit
ellipses to the galaxy image, varying the ellipticity and position
angle (while holding the ellipse centre fixed at the position of the
nucleus) so as to minimise residuals about the mode of the pixel
values around the ellipse perimeter. The semi-major axis is increased
by 15~per~cent each time (the increased number of pixels sampled at
each radius partly compensating for the reduced signal-to-noise in the
outermost pixels), until a level equal to the measured sky background
is reached.

The modal surface brightnesses from separate $B$ and $I$ analyses,
corrected for atmospheric (but not Galactic or internal) extinction,
are plotted in Figure~\ref{f:lumprof}. Also shown on this plot is the
{\em mean} surface brightness in \Halpha\ (with
$+24.0$~mag~arcsec$^{-2}$ being equivalent to an H$\alpha$ flux
density of $10^{-15}$~ergs~cm$^{-2}$ s$^{-1}$ arcsec$^{-2}$) computed
at each radius using the ellipse parameters from fitting to the
$I$-band image. The variation in ellipse inclination and position
angle as a function of radius in the $I$ band (less affected by dust
than $B$) has also been plotted in Figure~\ref{f:rotfig} for
comparison with those determined from the gas kinematics.

Blackman (1979) reported the presence of an extra linear outer
component, in addition to the inner bulge and exponential disk.
His reduced luminosity profiles are not directly comparable to
our radial surface brightness profiles, but Figure~\ref{f:lumprof}
does confirm the existence of a second exponential disk in the
range ($90\la r \la 200$ arcsec), with a longer scale length
than the inner exponential disk, which covers ($35 \la r \la
90$~arcsec). This outer exponential disk also coincides with
the transition region between the inner and outer \HI\ disks,
as well as the region of most rapid turnover in the rotation curve.

The integrated intensity out to the sky background yields apparent
magnitudes in $B$ and $I$ of $11.29\pm0.03$ and $9.47\pm0.02$
respectively.  After corrections for Galactic extinction
(A$_{B}=0.12$; Burstein \& Heiles 1984) and for extinction internal to
NGC~157 at an inclination of $(45\pm5)^{\circ}$ (Tully \& Fouqu\'{e}
1985), we arrive at corrected total magnitudes (in the RC3 convention)
of $B^{\rm o}_{T}=10.79\pm0.06$ and $I^{\rm
o}_{T}=9.28\pm0.05$. Adopting the same distance (20.9~Mpc) as used for
the total \HI\ mass determination yields an absolute magnitude
M$_{B}=-20.82\pm0.06$, and thus a total blue luminosity of
$\sim3.3\times10^{10}~L_{\odot}$. Our $B^{\rm o}_{T}$ value is
consistent with equivalent tabulations in RC3 and Tully (1988),
although the individual magnitudes and extinction corrections do
differ.  MFB found $I^{\rm o}_{T}$ some 0.3~mag fainter than us,
though the integrated $I$ magnitudes are in much better agreement, so
much of the difference lies in the (still contentious) corrections for
extinction.

The continuum-subtracted \Halpha\ data was similarly integrated out to
$D_{25}$, resulting in a total flux in the \Halpha\ line of $(2.1\pm0.2)
\times10^{-11}$~ergs~cm$^{-2}$~s$^{-1}$ ($\sim2.7\times10^8~L_{\odot}$
for the adopted distance). This includes a correction
for atmospheric extinction, as well as the 1.1~mag of internal
extinction applied by Kennicutt (1983) in his study of the star
formation rate (SFR) in normal disk galaxies. Using the same
conversion formulae from \Halpha\ luminosity to SFR as Kennicutt,
we estimate a total SFR for massive stars ($M\ga10~{\rm M}_{\odot}$)
of 1.6~M$_{\odot}$~yr$^{-1}$, and over all stellar masses (0.1~M$_{\odot}
<M<100$~M$_{\odot}$) to be $\sim10~{\rm M}_{\odot}$~yr$^{-1}$. Normalising
yields the following (distance-independent) quantities: $({\rm SFR}/
{\rm disk~area})=30~{\rm M}_{\odot}~{\rm pc}^{-2}~{\rm Gyr}^{-1}$;
$({\rm SFR}/L_{B})=0.3~{\rm M}_{\odot}~{\rm Gyr}^{-1}~L_
{\odot}^{-1}$; and $({\rm SFR}/M_{\rm HI})=1.3~{\rm Gyr}^{-1}$.
Consequently, although the absolute present-day SFR is not unusual for
its type (Sbc), relative to its size and mass, NGC~157 is nearly twice as
active in forming stars as galaxies of similar type and colour (Ryder 1993).

This enhancement of \Halpha\ emission is confirmed by the far-infrared
to \Halpha\ luminosity ratio, $L_{\rm FIR}/L_{{\rm H}\alpha}$. The
value of $\log L_{\rm FIR}$ obtained from $IRAS$ data after reducing to
the adopted distance is 10.34 (Young et~al. 1989); hence $\log L_{\rm FIR}
/ L_ {{\rm H}\alpha} = 1.9$, whereas for bright galaxies of a similar
colour, typical ratios are in the range $2.1 - 2.5$ (Zasov 1995). Note
that $\log L_{\rm FIR} / L_{B} = -0.18$, observed in NGC~157, is quite
normal for late-type spiral galaxies (Young et~al. 1989; Zasov 1995),
in contrast to the normalised H$\alpha$ star formation rates.
Using the relationship between star
formation rate and far-infrared luminosity (${\rm SFR} \sim 2.5 \times
10^{10} L_{\rm FIR}/L_{\odot}$; Zasov 1995), we obtain ${\rm SFR} \sim 5
M_ {\odot}$~yr$^{-1}$. No \Halpha\ emission was detected in the
extended \HI\ disk, to a $3\sigma$ limit of
$3\times10^{-16}$~ergs~cm$^{-2}$~s$^{-1}$~arcsec$^{-2}$, implying that
the massive star formation rate ($M\ga10$~M$_{\odot}$) there cannot
exceed $\sim2$~M$_{\odot}$~pc$^{-2}$~Gyr$^{-1}$.

\section[]{Discussion}\label{s:disc}

\subsection[]{A superbubble in NGC~157?}

Given that one of the primary motivations for this study was the
possible existence of a major \HI\ superbubble in the disk of NGC~157,
how convincing is the new evidence for or against such a feature?
Normally, the strongest evidence for such superbubbles
comes from the localised distortions they introduce to the global
velocity field (Ryder et~al. 1995), and yet the isovelocity contours in
Figure~\ref{f:nah1vf} (as well as the higher resolution images) exhibit a
remarkable degree of symmetry about both the major and minor axes.
Examination of the position-velocity diagrams in this region also
indicates no unusual profile splitting or deviations attributable to
the action of an expanding superbubble, although the fact that the
projected rotation velocity is changing rapidly in this particular
region complicates such an analysis.

\begin{figure*}
\vspace{22cm}
\includegraphics{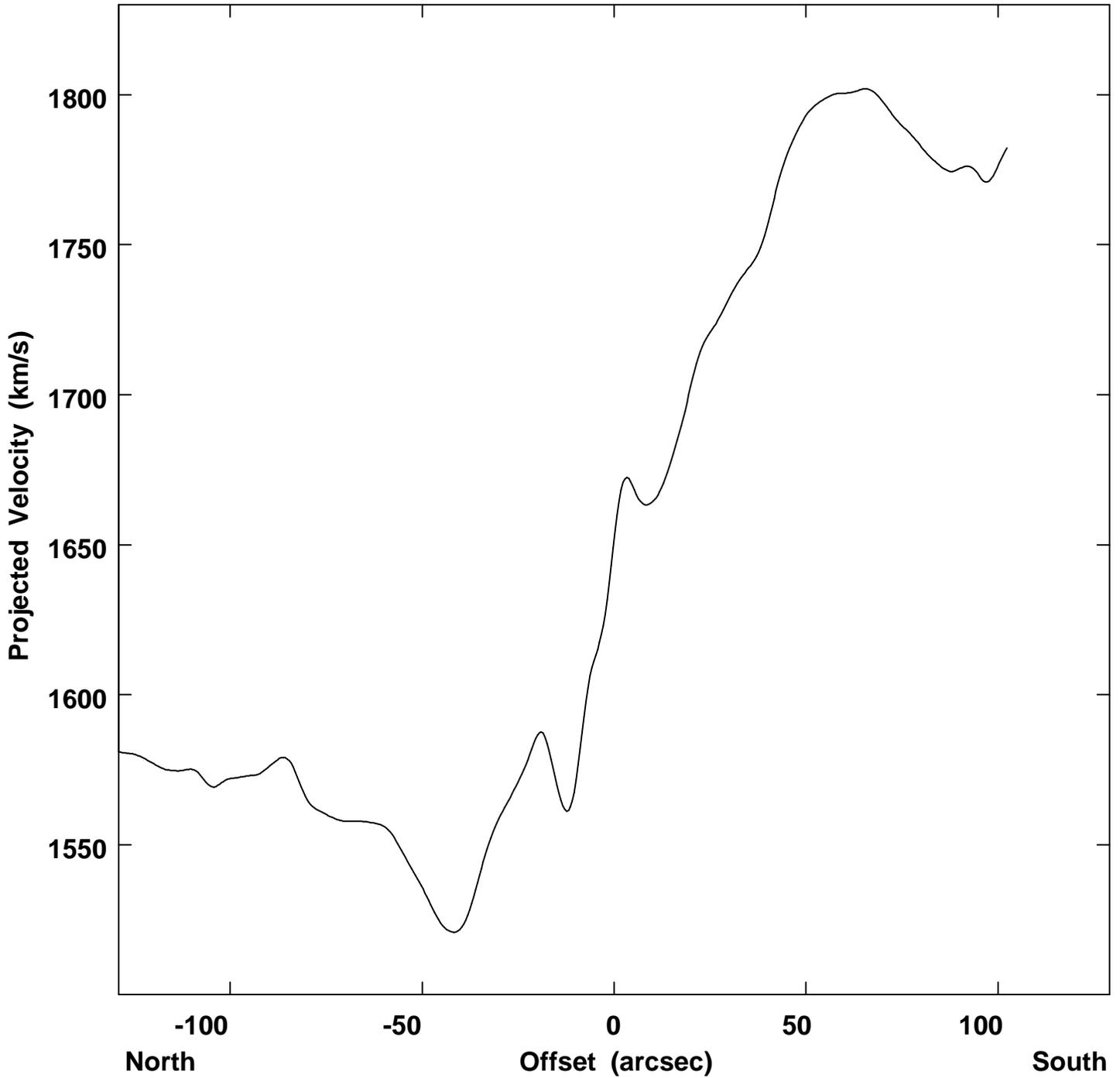}
 \caption{A cut through the uniform-weighted \HI\ cube at a position angle
of $10^{\circ}$, for comparison with Figure~3 of Zasov \& Kyazumov (1981).}
\label{f:n157cut}
\end{figure*}

How then are we to explain the abrupt change in velocity across the
northeast ring seen by ZK in \Halpha? We have made cuts through the
uniform-weighted \HI~cube at similar position angles (e.g.,
$\theta=10^{\circ}$; Figure~\ref{f:n157cut}), and find that beyond
45~arcsec along the northeastern (approaching) axis, there is indeed a
drop of almost 40~km~s$^{-1}$ in velocity which, when deprojected by
the same inclination ($i=31^{\circ}$) used by ZK and Burbidge
et~al. (1961), results in a total drop of $\sim80$~km~s$^{-1}$, as
reported by ZK. The \HI~cut however shows a similar drop on the
opposite axis beyond 60~arcsec, by which point ZK were unable to
detect any more optical emission. Thus, the velocity discontinuities
reported by ZK are real, but are caused by something other than the
action of an \HI~superbubble.

\subsection[]{The bar in NGC~157}\label{s:bar}

Surface photometry in the near infrared ($JHK$) bands has recently
been published by Elmegreen et al. (1996a, b), who make the case for a
bar component, with a flat luminosity profile going out to one-quarter
of $r_{25}$. Sempere \& Rozas (1997) have carried out numerical
simulations of the interstellar medium in NGC~157 under the action of
a bar-driven spiral density wave, and predicted the locations of
several resonances accordingly.

\begin{figure*}
\vspace{22cm}
\includegraphics{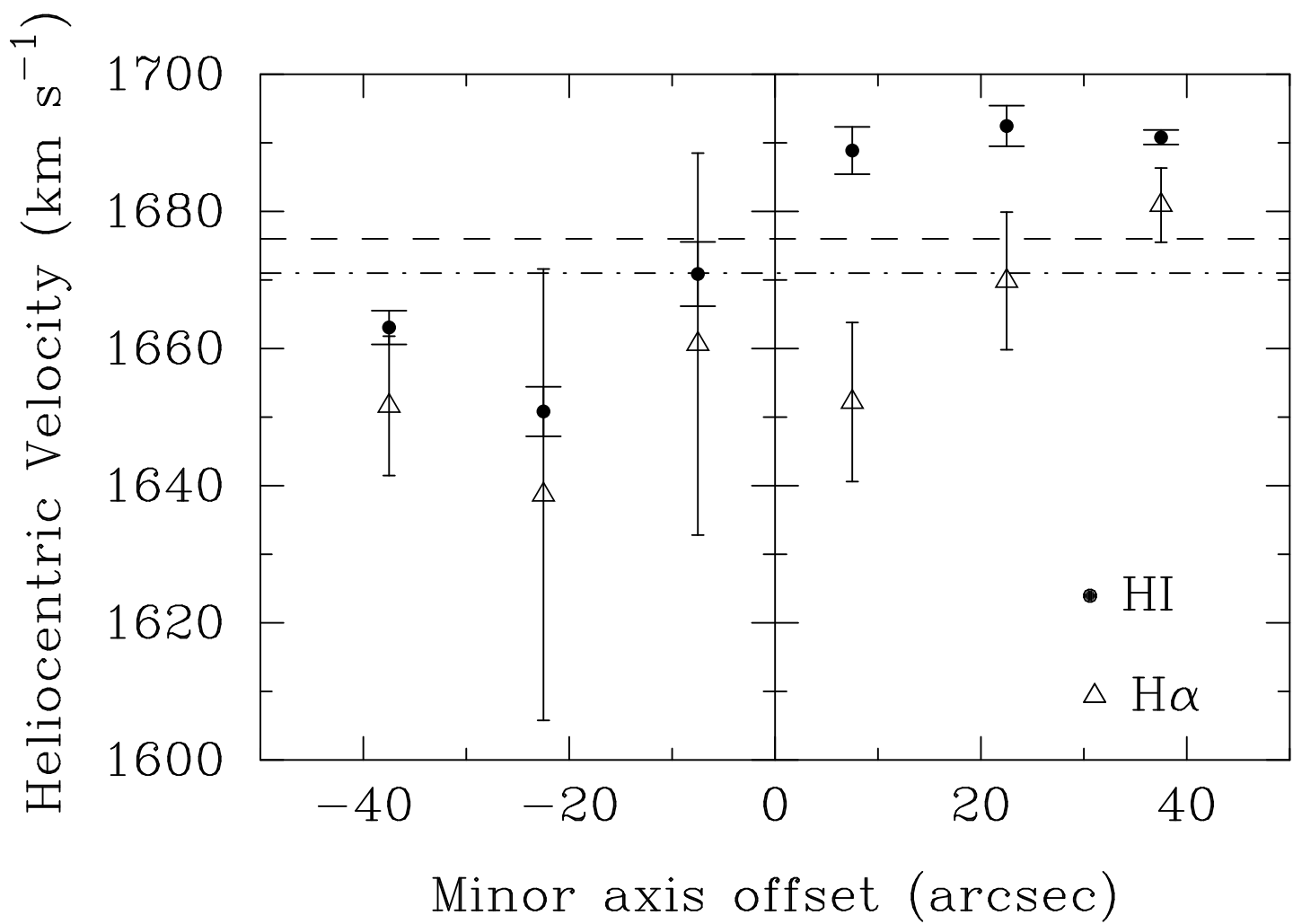}
 \caption{Cut along the minor axes of the \Halpha\ and \HI\
velocity fields, with data binned into 15~arcsec intervals.
The solid points and fat error bars are the mean and r.m.s.~velocities
from the uniform-weighted \Halpha\ data, while the open triangles
and thin error bars are from the \Halpha\ data. The dashed line is
the systemic velocity derived from fitting to the \HI\ velocity
field, while the dash-dotted line is that from the \Halpha\
velocity field.}
 \label{f:minaxvp}
\end{figure*}

In an effort to see how the kinematics may be influenced by the action
of a bar, we have sampled both the \Halpha\ and \HI\ velocity fields
along the kinematic minor axis ($\theta=310^{\circ}$), and binned the
data into 15~arsec intervals, to match the resolution of the \HI\
data. As Figure~\ref{f:minaxvp} shows, both the neutral and ionised
gas components exhibit almost sinusoidal variations about the systemic
velocity along the minor axis. Assuming the spiral arms to be trailing,
then the observed rotation pattern implies that the southeast side
of the disk is closest to us. Consequently, the negative velocity
residuals in Figure~\ref{f:minaxvp} along the southeast half of the
minor axis indicate gas flowing towards the observer, and therefore
an {\em outflow}, rather than the gas inflow more commonly assumed
to be associated with the action of a bar. Thus the nuclear kinematics
appear to be dominated by the presence of a mild starburst, rather
than any associated bar.


\subsection[]{The turnover in the rotation curve of NGC~157}\label{s:tover}

The extent and abruptness of the decline in rotation velocity in
NGC~157 would, if proved, make it one of the most unusual rotation
curves yet seen.  The prevalence of flat rotation curves as seen in
the ionised gas (Rubin, Ford \& Thonnard 1978) and in the
extended \HI\ (Bosma 1978) has been taken as one of the strongest
arguments in favour of the existence of a dark matter halo around all
spiral galaxies. To date, only a handful of galaxies are observed to
have truly declining \HI~rotation curves, consistent with having
reached the edge of the {\em total\/} mass distribution; these include
NGC~2683 and NGC~3521 (Casertano \& van~Gorkom 1991), NGC~7793
(Carignan \& Puche 1990), NGC~1365 (J\"{o}rs\"{a}ter \& van~Moorsel
1995), and NGC~4244 (Olling 1996). However, NGC~157 is somewhat more
unique, in the sense that the decline in rotation velocity
is both rapid and severe, and yet the outermost part of the rotation
curve is still relatively flat, indicating that the mass distribution
continues well beyond the last measured point.

There are a few galaxies where optical observations have also revealed
a decline in the line-of-sight velocities; for example NGC~4303,
NGC~4321 (Distefano et~al. 1990), NGC 4254, and NGC~4536 (Sperandio
et~al. 1995; Afanasiev et~al. 1992). However, none of these cases show
a decline as abrupt as we have found in NGC~157. More importantly, in
all cases cited, the decline is seen only on one side of the nucleus,
and often the decline is not visible in the \HI~rotation curve (e.g.,
NGC~4254; Distefano et~al. 1990). In contrast, our \HI~data are in
good agreement with ZK's measurements, and while the optical decline
is seen on only one side of the galaxy, the decline is seen at the
same radius on both sides of the galaxy in \HI.

\begin{figure*}
\vspace{22cm}
\includegraphics{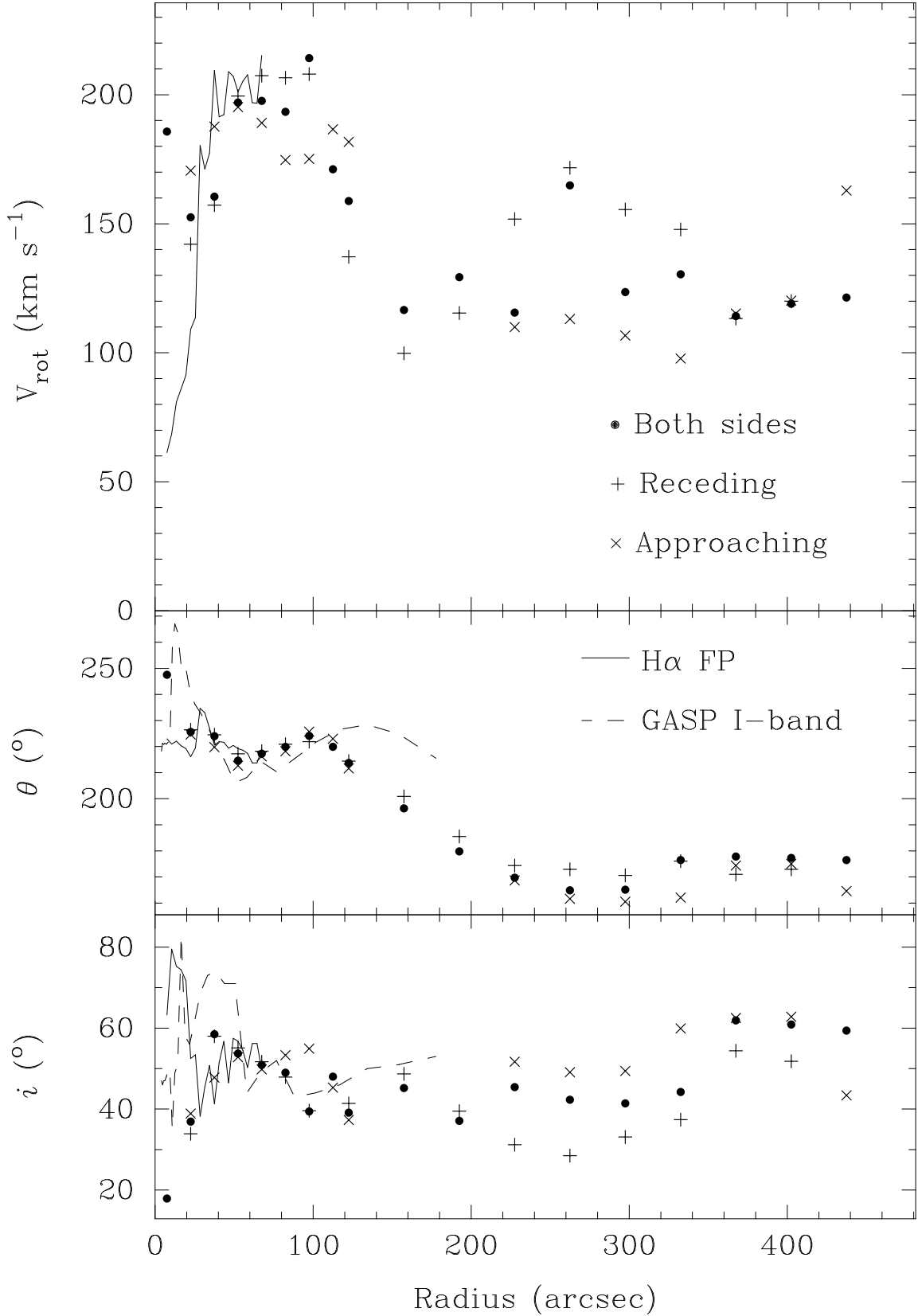}
 \caption{Rotation curve and variation of ring inclination and kinematic
line-of-nodes from ROTCUR modelling of the uniform ($r<120$~arcsec) and
natural-weighted ($r>120$~arcsec) velocity fields. Points indicate the
fits to the observed \HI\ velocity field, the solid line is the equivalent
result from the \Halpha\ velocity field, and the dashed line traces the
orientation parameters from surface photometry of the $I$-band image.
For the \HI~velocity field, separate analyses have been carried out on
the entire disk ($\bullet$), as well as for just the receding (+) and
approaching ($\times$) halves. Error bars in the formal fitting are much
smaller than the differences between the separate halves.}
 \label{f:rotfig}
\end{figure*}

\begin{figure*}
\vspace{22cm}
\includegraphics{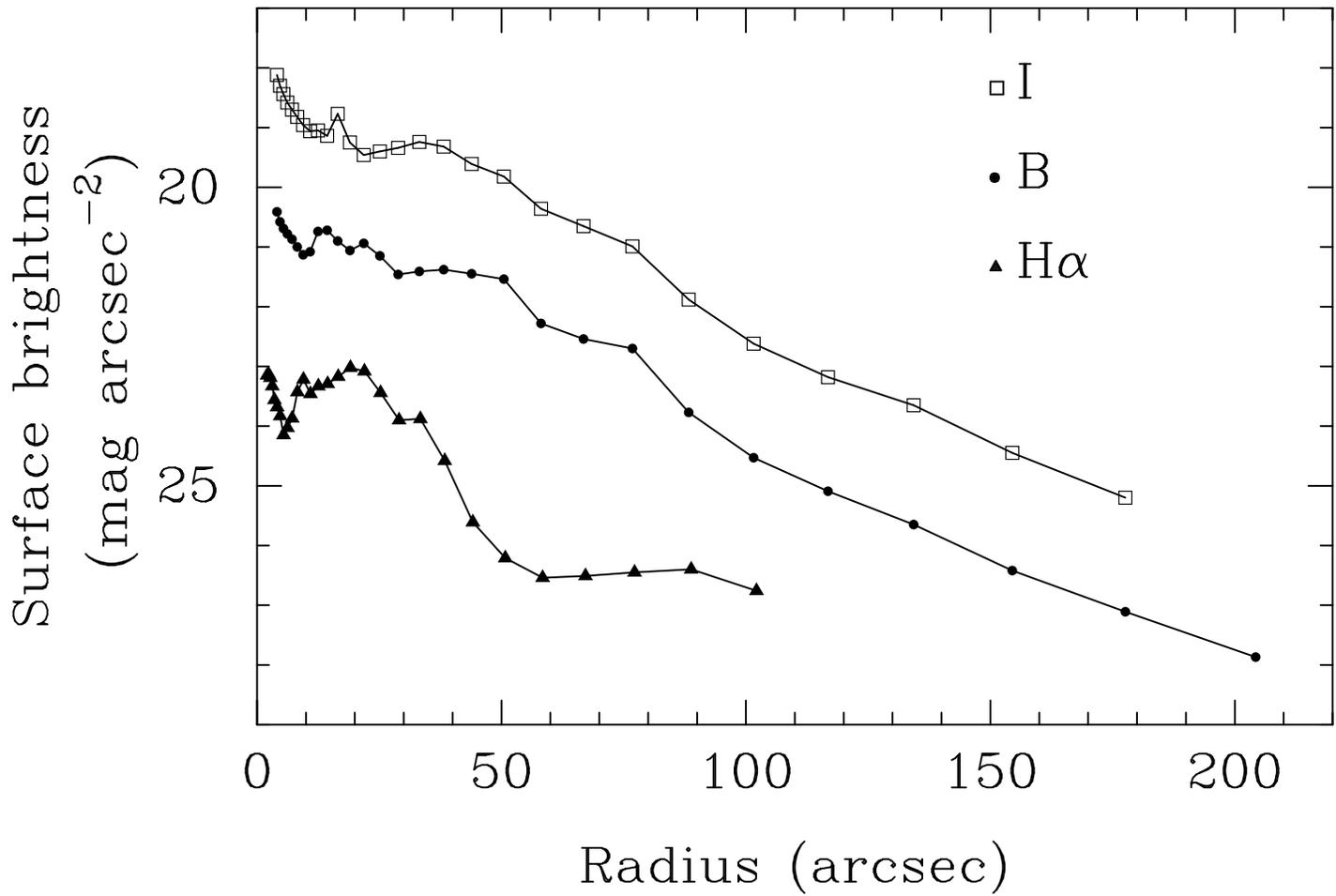}
 \caption{Radial surface brightness profiles for NGC~157 from ellipse fitting
to the $B$, $I$, and \Halpha\ images. The disk orientation parameters found
from the $I$-band fitting are shown separately in
Figure~\protect{\ref{f:rotfig}}.}
 \label{f:lumprof}
\end{figure*}

\begin{figure*}
\vspace{22cm}
\includegraphics{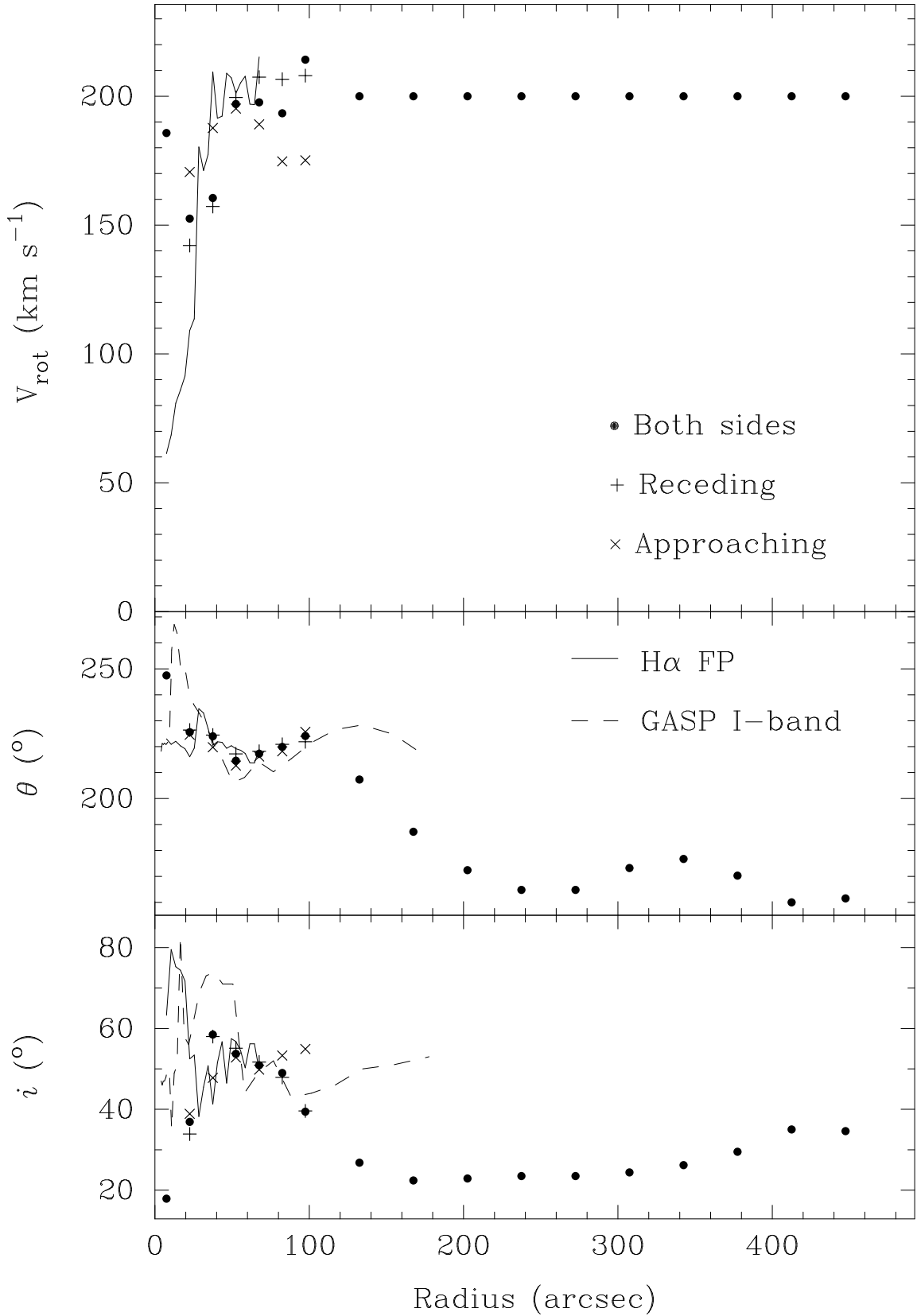}
 \caption{As for Figure~\protect{\ref{f:rotfig}}, except that the rotation
velocity is held fixed at 200~km~s$^{-1}$ for $r>100$~arcsec.}
 \label{f:v200}
\end{figure*}

Before attempting to model the mass distribution, we must be
confident that Figure~\ref{f:rotfig} is a fair and accurate
representation of the disk rotation and orientation parameters in
NGC~157. It could be argued that a more likely scenario is that
the rotation curve stays flat, and the change in projected rotation
velocity seen in the outer disk is entirely due to variations in
$\theta$ and $i$. In Figure~\ref{f:v200}, we have fixed $V_{\rm rot}$
at 200~\kms\ (close to the maximum value attained in the inner
disk) for $r>100$~arcsec, and used the ROTCUR routine to again find
the combination of $\theta$ and $i$ that minimises the circular
velocity residuals. These residuals average 9.6~km~s$^{-1}$, only
slightly larger than the 7.9~km~s$^{-1}$ residuals attained with the
free fit shown in Figure~\ref{f:rotfig}.

The warp behaviour of the line-of-nodes is almost unchanged, but the
inclination is forced to drop as low as $23^{\circ}$ in order to
sustain such a high rotation velocity. Requiring a constant rotation
velocity of only 180 or 160~km~s$^{-1}$ results in a minimum
inclination of $25^{\circ}$ and $30^{\circ}$ respectively. The field
galaxy calibration of the Tully-Fisher relation (Pierce \& Tully 1992)
for M$_{B}=-20.82\pm0.06$ implies a still higher maximum rotational
velocity of $234^{+137}_{-86}$~\kms.  To match this velocity with a
flat rotation curve would require $i\sim35^{\circ}$ in the {\em
inner\/} disk, dropping to as low as $i\sim25^{\circ}$ at
$r=400$~arcsec. Thus, it appears that the outer \HI~velocity is
unusually low, rather than the inner \HI~velocity peak being
anomalously high, for a galaxy of this magnitude. These low
inclinations are consistent with the axis ratios of the outer
\HI~contours (Figure~\ref{f:nah1onb}), but countering this is the fact
that the $I$-band surface photometry does not support such low
inclinations in the region $90~<r<180$~arcsec. At these radii, the
spiral arms do not strongly disturb the isophotes, and to account for
the difference by intrinsic ellipticity of the disk requires
$e\sim0.2$, an extreme value ({\em cf.} 0.05; Rix \& Zaritsky
1995). Thus, unless the gas and the stars are somehow `decoupled'
(e.g., Section~\ref{s:trunc}), and orbit in totally different planes,
the rotation curve must decline significantly in this region.

One other possibility is that gas orbits in the outer disk of NGC~157
are intrinsically {\em elliptical}, violating one of the principal
assumptions of the ROTCUR analysis, and casting doubt on the derived
inclinations. Non-circular gas motions are often seen in the central
regions of galaxies, usually arising through the influence of a bar.
These effects are usually restricted to the region near the bar
(though there are exceptions in cases where a resonance is set up
between the orbital epicyclic frequency and the bar pattern speed:
Ryder et~al. 1996); outside of this, streaming motions along the
spiral arms dominate perturbations from uniform circular motion. In
NGC~157, the bar component proposed by Elmegreen et~al. (1996a, b) and
by Sempere \& Rozas (1997) is a small-scale feature, extending to a
radius of only $\sim0.25r_{25}$ (30~arcsec), so it can be discounted as
the source of the velocity peak in the rotation curve (Section~\ref{s:bar}).

To check the effects of elliptical orbits at large radii, we computed
a few model velocity fields and analysed them with ROTCUR. The
elliptical orbits were not computed from an assumed potential, but as
Lissajous figures, the sum of two simple harmonic oscillators. We
compared orbit ellipticities ($e=(a-b)/a$) of $e=0$ and $e=0.5$, for a
flat $\langle v \rangle = 200$~\kms\ rotation curve and a declining
curve, computed from the Brandt (1960) formula for $R_{\rm
max}=80$~arcsec, $V_{\rm max}=200$~\kms and $n=10$. The calculations
were done for disks inclined at $i=45^{\circ}$ and $60^{\circ}$ to the
line of sight. For this small grid of models, the shapes of the
outer-disk isovelocity contours remained almost unchanged by taking
the orbits to be oval, but the amplitude of the radial velocity
variations was increased. ROTCUR correctly fitted the inclination and
position angles, but for the elliptical-orbit velocity fields it
returned circular velocities $\sim25$\% higher than the input
value. For the $i=45^{\circ}$ declining rotation curve case, ROTCUR
had difficulty in fitting the inclination, and did not give consistent
results from ring to ring, or for different initial
parameters. However, $e=0.5$ is an extreme value, and given the lack
of any effect in the other test cases, we think that the velocity
field of NGC~157 can be well modeled by circular orbits.

Finally, we note that although he had no knowledge of the extended
\HI\ disk (nor its unusual kinematics) in NGC~157, Blackman (1979)
did investigate the predictions of the spiral density wave theory
by {\em assuming\/} a declining rotation curve outside of $r_{25}$,
of the form popularized by Brandt (1960). He found that such a rotation
curve yielded a theoretical spiral pattern in much better agreement with
the actual spiral arm pattern than did a flat rotation curve in the outer
regions. Although this approach involved many simplifications, it does
provide yet another clue in favour of the declining rotation curve.

In summary, the velocity field is probably not strongly affected by
peculiar motions, but the difficulties in separating the effects of
inclination and rotation velocity remain, and all we can really say is
that the actual rotation curve behaviour almost certainly lies
somewhere between the two extremes shown in Figures~\ref{f:rotfig} and
\ref{f:v200}. We can then use these to place upper and lower bounds on
the various mass models.

\subsection[]{Mass Models of NGC~157}

In this section, we consider some scenarios that could account for
the peculiar velocity field/rotation curve in NGC~157.

\subsubsection{`Missing' Dark Matter?}

\begin{figure*}
\vspace{22cm}
\includegraphics{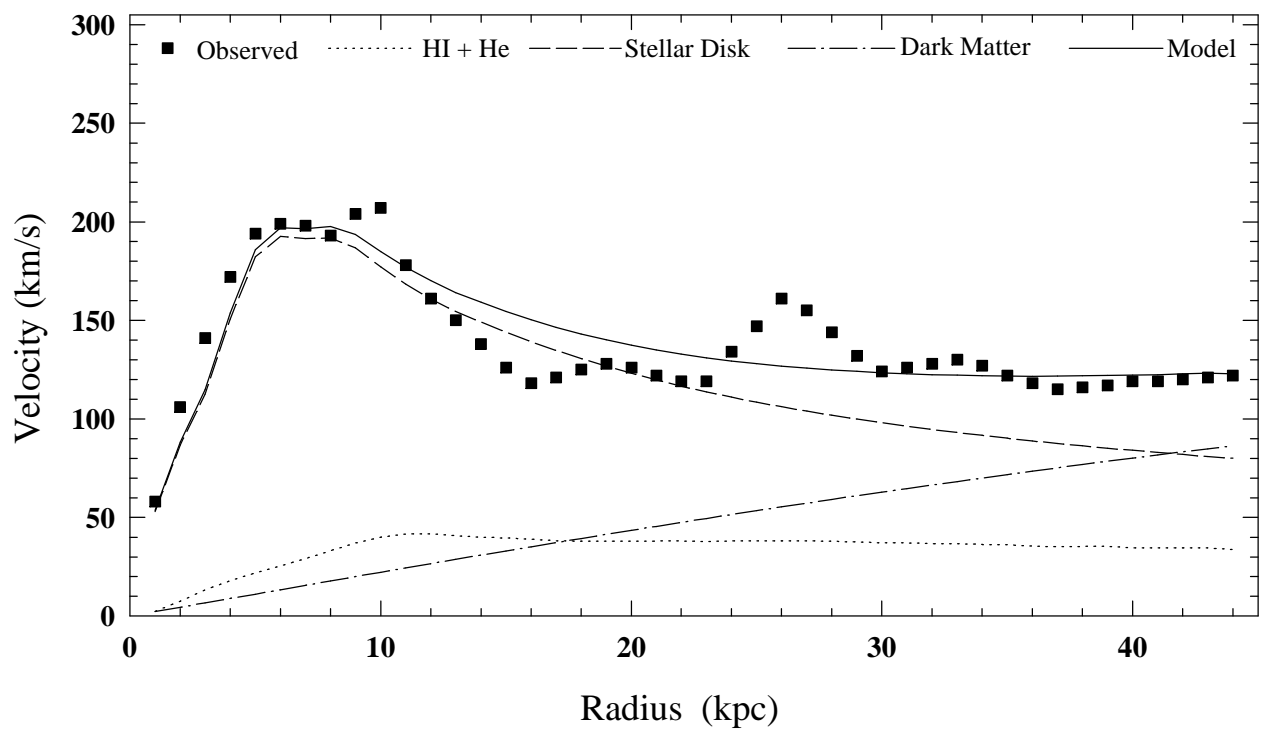}
 \caption{Plot of the relative contributions to the declining rotation
 curve (squares) of NGC~157 from (dotted line) atomic gas,
 (dashed line) stars, and (dot-dash line) a dark matter halo,
 with the sum of these three shown as the solid line.}
 \label{f:wwfig}
\end{figure*}

We begin with the standard `maximum-disk' model for fitting the
declining and flat rotation curves, using the procedure described in
Walsh, Staveley-Smith, \& Oosterloo (1997). The relative contributions
to the overall rotation due to the stellar component, the gas surface
density ($\sigma_{\rm HI}$ multiplied by 1.3 to account for the
presence of He), and a pseudo-isothermal halo have been calculated
using a least-squares fit to the observed rotation curve. A summary of
the best fitting parameters is given in Table~\ref{t:wwpars}, and the
fit to the rotation curve with the turnover is plotted in
Figure~\ref{f:wwfig}.  The mass of the gaseous disk is
$9.62\times10^{9}$~M$_{\odot}$, and the stellar disk has
$1.92\times10^{11}$~M$_{\odot}$ (with a stellar $M/L_{I}$ of 1.4). The
gas mass-to-stellar mass ratio is therefore quite normal at 5\%.

\begin{table*}
\begin{minipage}{120mm}
 \caption{Mass model properties for NGC~157.}
 \label{t:wwpars}
 \begin{tabular}{lccc}
            & Declining R.C. & \multicolumn{2}{c}{Flat R.C.} \\
Parameter   &                & $V_{\rm max}=170$~km~s$^{-1}$ &
$V_{\rm max}=200$~km~s$^{-1}$ \\
\hline
$M/L_{I}$ (M$_{\odot}$/L$_{\odot}$) & 1.4    & 0.84   & 0.94 \\
$\rho_{0}$ (M$_{\odot}$~pc$^{-3}$)  & 0.0019 & 0.0071 & 0.014 \\
$r_{\rm c}$ (kpc)                   & 15     & 9.5    & 7.6 \\
$M_{\rm dark}$ ($10^{11}$~M$_{\odot}$)$^{a}$ & 2.1 & 3.4 & 4.3 \\
\end{tabular}
\medskip
~~\\
$^{a}$Mass of dark matter contained in halo within $r<44$~kpc.
\end{minipage}
\end{table*}

In the case of the declining rotation curve, the dark-to-luminous
matter ratio at the last measured point is unusually low, at almost
1:1. This is almost identical to the value found in NGC~7793 (Carignan
\& Puche 1990), and only the massive (and distant) galaxy NGC~801 in
the sample of BvW has a lower value. Based on the empirical relations
of Kormendy (1990), we would have expected a central density
$\rho_{0}\sim0.004$~M$_{\odot}$~pc$^{-3}$ and a core radius $r_{\rm
c}\sim19$~kpc for a galaxy with the blue luminosity of NGC~157.  This
is quite a large core radius, and once again, only the most luminous
($M_{B}>-21$) galaxies in the BvW sample have larger core radii and
lower central densities.

Even for the case of a flat rotation curve, the dark-to-luminous mass
ratio still only approaches 2 at the last measured point, though such
a value is more in keeping with that measured in galaxies of similar
$V_{\rm max}$ and Hubble type (BvW). In summary, a standard gas $+$
stellar disk $+$ dark matter halo model points toward NGC~157 having a
low dark matter content, though not unprecedentedly so.

This raises the question of how the galaxy could have come to have a
low dark matter content. Halo stripping in a cluster environment is
easy enough to understand (Whitmore et~al. 1988), but as noted in the
introduction, NGC~157 is rather isolated. Casertano \& van~Gorkom
(1991) found that bright (massive), compact (i.e. short disk scale
length) galaxies are more likely to show declining rotation curves
(usually by no more than 30\% of their peak velocity), though BvW
disputes the scale length dependence. NGC~157 possesses both a fast
rotating and relatively compact disk, with an $I$-band scale length of
3.4~kpc; according to Casertano \& van~Gorkom (1991), such galaxies
will have a steeper potential well, into which dark matter is less
likely to fall. Thus, if the dark matter content of NGC~157 is as low
as the decline in the rotation curve suggests, then it presumably has
always been low, reflecting the circumstances of its birth.

\subsubsection{Truncated or Decoupled Disks}\label{s:trunc}

Casertano (1983) showed how a truncated exponential disk could give
rise to an abrupt decline in the rotation curve just beyond the
truncation radius. In particular, he was trying to account for a
sudden drop in the rotation curve of the edge-on spiral galaxy
NGC~5907, but photometry of such galaxies (e.g., van~der~Kruit \&
Searle 1981) had already made a strong case for truncations in the
light distribution.

We have used a modified version of the disk-fitting procedure from the
previous section in order to test whether a truncated disk could
account for some (or all) of the drop in rotation velocity.  Based
partly on the findings of Blackman (1979), confirmed by our
Figure~\ref{f:lumprof}, that NGC~157 appears to have a second, outer
exponential disk, we include 2 separate disk components in our models;
one truncated at some radius $R_{\rm t}$, and a second, less massive
untruncated disk to account for the velocity of rotation at large
radii.  We therefore found it necessary to model the bulge component
separately, using a King profile.  The scale length of the inner disk
was taken to be 3.4~kpc, derived from a linear fit to the $I$-band
profile of Figure~\ref{f:lumprof} in the range $35<r<100$~arcsec. This
disk was truncated at a radius $R_{\rm t}$, with the best fit obtained
for $R_{\rm t} = 6 - 6.5$~kpc, i.e., $\sim2$ disk scale lengths. To
`soften' the sharp fall of the density at the edge, a non-zero scale
length (1~kpc) was introduced for $R \oldge R_{\rm t}$.  The
parameters of the untruncated exponential disk, which dominates the
density at $R\oldge R_{\rm t}$ were then varied to give the best fit
to the observed rotation curve. The best fit value of the scale length
was found to be 4.7~kpc, which is slightly larger than was found from
the $I$-band surface photometry (Section~\ref{s:sphot}). The
asymptotic velocity of rotation of the model halo is only 110~\kms,
which is about half of the maximum velocity of rotation of the disk.

\begin{figure*}
\vspace{22cm}
\includegraphics{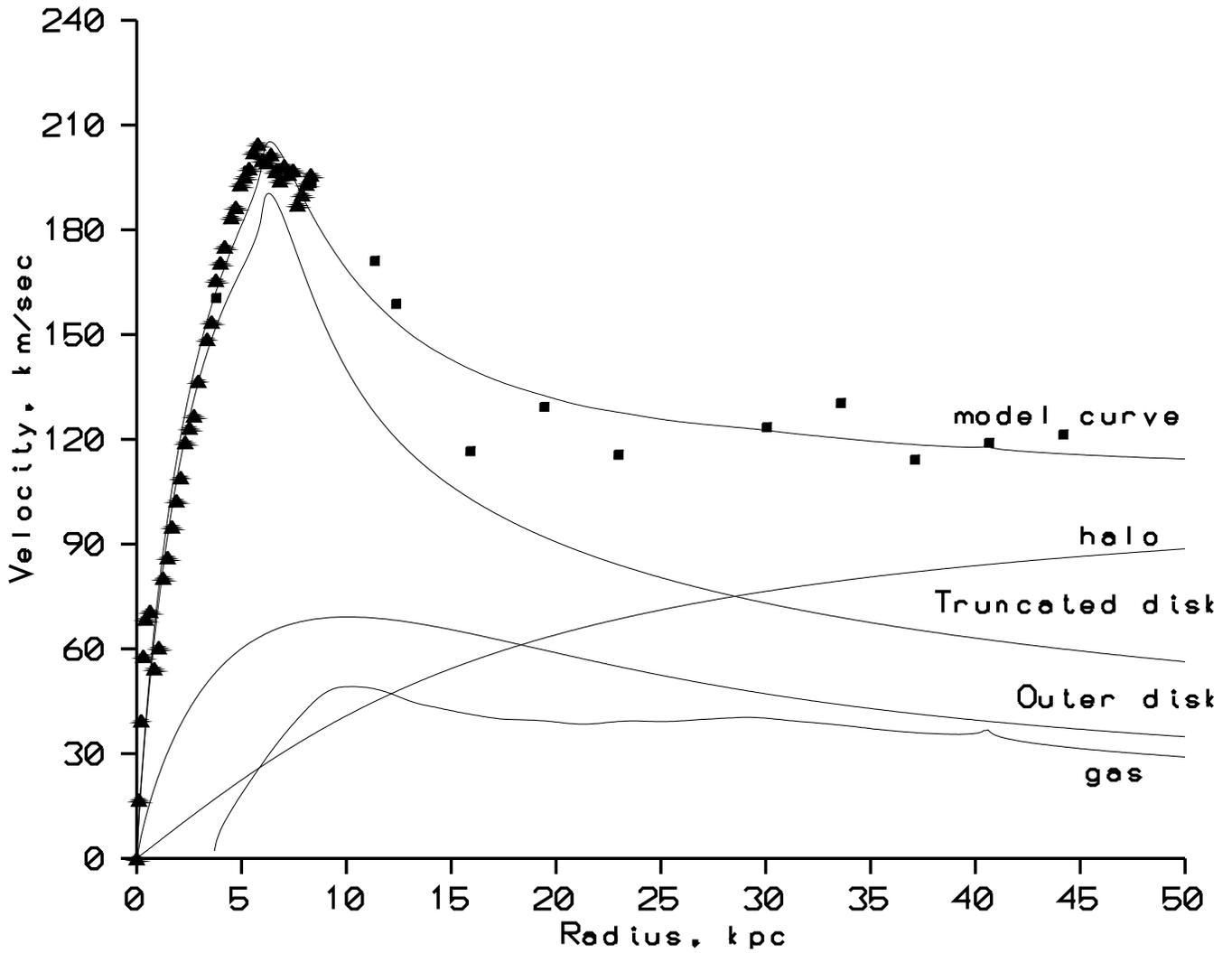}
 \caption{Plot of the relative contributions to the declining rotation
 curve (solid symbols) of NGC~157 from an exponential disk truncated
 at about 2 scale lengths, an outer exponential disk, atomic gas, stars, and a
 dark matter halo, as well as their sum.}
 \label{f:avz}
\end{figure*}

The rotation due to each component, plus the total compared with the
observed declining rotation curve, is plotted in Figure~\ref{f:avz}.
The masses of individual components within three important radii are
given in Table~\ref{t:avzpars}: $r = 12.5$~kpc, which corresponds to
the optical radius $r_{25}$ of this galaxy; $r = 25$~kpc, where the
halo begins to dominate; and $r=50$~kpc, just outside the last
measured point. The dark-to-luminous mass ratio at these points grows,
from 0.13, to 0.51, and finally 1.56 respectively. Thus, although the
truncated disk model can successfully account for the rapid drop in
rotation velocity beyond 5~kpc, the outer exponential disk and gas
components together are not sufficient to sustain the rotation curve,
and a dark matter halo (albeit a minimal one) is still
required. Taking into account the total blue luminosity of the galaxy
(Section~\ref{s:sphot}), we obtain an unusually low $M/L_B \sim 0.6$
(solar units) within $r_{25}$, which further supports the conclusion
that there is no significant dark halo within the luminous disk.

\begin{table*}
\begin{minipage}{120mm}
 \caption{Truncated disk model properties within given radius for NGC~157.}
 \label{t:avzpars}
 \begin{tabular}{lccc}
Model component & \multicolumn{3}{c}{Mass within $R$ ($10^{9}$~M$_{\odot}$)} \\
                &  $R=12.5$~kpc & $R=25$~kpc & $R=50$~kpc \\
\hline
Truncated disk  &       36.0    &     36.0   &    36.0    \\
Outer disk      &       10.0    &     13.0   &    13.4    \\
Bulge           &       $<1$    &     $<1$   &    $<1$    \\
Gas (\HI\ $+$ He) &      4.5    &      7.1   &     8.6    \\
Halo            &        6.7    &     29.0   &    91.0    \\
Total           &       57.0    &     85.0   &   149.0    \\
\end{tabular}
\medskip
\end{minipage}
\end{table*}

Truncated disks are seldom evident in radial surface brightness
profiles of low-inclination galaxies, most likely because of
azimuthal averaging in the analysis smearing out any asymmetric
edges in the younger stellar populations (van~der~Kruit 1988).
Radial cuts across the $B$-band image of NGC~157 shown in
Figure~\ref{f:nah1onb} give the impression of a more abrupt
drop in luminosity towards the northwest edge of the disk
than elsewhere, though this is at least partly a matter of
contrast with the outer edge of the northern spiral arm.

Truncated disks and warps might be related, since Sparke (1984) showed
that a sharply truncated edge to the disk could be the crucial factor
in sustaining vertical oscillation modes (i.e., a warp) in the
presence of a halo. Of course, the disk referred to in this case is
the gaseous disk, whereas it is the stellar disk that is truncated in
our model; furthermore, Sparke's model produces warps {\em interior\/} to
the truncation. Nevertheless, the forces that give rise to one
phenomenon (a warp or a truncated disk) could equally well be driving
the other.

We wish to stress that the phenomenon we are interested in here is
that of a truncated {\em inner\/} disk, rather than an actual edge to
the total mass distribution (as might be produced say, by ram-pressure
stripping: Gunn \& Gott 1972; Kritsuk 1983) or to just the \HI\ disk
(e.g., photoionisation by the extragalactic UV radiation field:
Maloney 1993; Dove \& Shull 1994). All these indications of a
truncated inner disk lead us to reconsider our other data under the
hypothesis that NGC~157 is a loosely-coupled (or perhaps even merged)
{\em pair\/} of disks. The suddenness of the decline in projected rotational
velocity, the coincidence of this with the optical edge of the disk,
and the apparent discrepancy between optical and kinematic
inclinations (if we require a flat rotation curve), all lend some
support to this idea.

The total \HI\ spectrum gives conflicting evidence. The (slightly)
different systemic velocities of the inner and outer \HI~disks would
favour this scenario, but the roughly equal gas masses in the two
components is a contra-indication, as the merging of equal-mass disks
is expected to result in an elliptical galaxy (e.g., Mihos \&
Hernquist 1996, and references therein). Indeed, even minor accretions
of mass are expected to cause significant heating of galaxy disks
(T\'{o}th \& Ostriker 1992; Quinn, Hernquist, \& Fullagar 1993;
Walker, Mihos, \& Hernquist 1996).  On the other hand, the \HI~gas in
the merging system NGC~520 is still recognisably `disky', with a
rather smooth velocity field, despite showing tidal tails, strong
distortion of the optical light, and other signs of a disk-disk merger
(Hibbard 1995). Similarly, NGC~4826 is thought to have survived as a
disk system after a major addition of mass (Braun et~al. 1994). This
galaxy has an outer gas disk that counter-rotates with respect to the
inner gas and stars, the two disks having identical kinematic
orientation parameters. Braun et~al. propose that it has formed either
by a slow accretion of retrograde gas, or the merger of two disks of
opposite spin, one much richer in gas than the other. While the
chances of such an orderly merger must be extremely slim, other
examples of counter-rotating disk components are appearing (e.g., Jore
et~al. 1996), so the possibility that NGC~157 is also a multiple
system cannot readily be dismissed.

\section{Conclusions}\label{s:conc}

The new radio and optical observations of NGC~157 presented in this
paper put this object among a small number of spiral galaxies whose
photometric and dynamical properties have been studied in detail over
a large range of radial distances - from less than one~kpc from the
nucleus up to nearly 50~kpc, or about 14~optical scale lengths.
Although many properties of this galaxy are rather common (grand
design spiral structure, the total mass of luminous components,
the maximum velocity of rotation, etc.), in some other respects,
NGC~157 is a most peculiar system:
\begin{enumerate}
\item NGC~157 possesses a highly extended \HI\ disk which is
mildly warped and stretches far beyond the optical extent of the
galaxy.
\item The \HI\ in the inner galaxy is not only concentrated in the
spiral arms well defined by \HII~regions, but also shows remarkable
honeycomb-like structure in the interarm regions, which in turn gives
a strong impression of `superbubbles' in the disk.
The azimuthally-averaged surface density of \HI\ drops sharply in the
outer part of the optical disk: it decreases by more than a factor of
three in the narrow range $80-120$~arcsec ($8 - 12$~kpc).
\item The combined velocity field of both ionised and neutral hydrogen
implies an unusual shape for the rotation curve, which begins to
decline steeply in the outer part of the optical disk and continues to
drop until the velocity of rotation reaches almost half of its
maximum value.  This points to the presence of a dark matter halo, the
relative mass of which is small when compared to the mass of the
stellar disk; a low $M/L_{B}$ ratio $(\sim 2)$ supports this
conclusion.  It is argued that NGC ~157 is the only known galaxy for
which a low-mass halo may be claimed with some confidence.
\end{enumerate}

We have outlined various circumstances under which NGC~157 may have
attained these features, including an unusually steep potential well,
the combination of truncated and extended exponential disks,
or possibly a merger of two equal-mass gas disks. Although we are not
yet able to distinguish which (if any) of these scenarios is in
operation in the case of NGC~157, we do point out that the distinctive
`shoulders' on the global \HI\ profile may be useful for identifying
potential analogs of NGC~157, with large and abrupt declines in
their rotation curves.

\section*{Acknowledgments}

We wish to thank R.~Buta, D.~Crocker, and M.~Lewis for making it
possible to acquire the optical images of NGC~157, and J.~Boulesteix,
S.~Dodonov, and the {\em Laboratory of Spectroscopy and Photometry of
Extragalactic Objects\/} of the Special Astrophysical Observatory for
assistance with the Fabry-Perot observations and reductions. We
acknowledge useful discussions with J.~Bland-Hawthorn, R.~Ekers,
J.~Higdon, P.~Sackett, L.~Sparke, T.~Oosterloo, and
D.~Westpfahl. S.~D.~R. acknowledges support from EPSCoR grant
EHR-9108761, and the receipt of a UNSW Vice-Chancellor's Postdoctoral
Research Fellowship. V.~J.~M. thanks the Smithsonian Institution for a
Predoctoral Fellowship, and the University of Wollongong for an
Australian Postgraduate Research Award. A.~V.~Z. and O.~S. ackowledge
the Russian Fund for Basic Research for partial support of this work
(grant 96-02-19197). The National Radio Astronomy Observatory is a
facility of the National Science Foundation operated under cooperative
agreement by Associated Universities, Inc. We have made extensive use
of the Lyon-Meudon Extragalactic Database (LEDA, supplied by the LEDA
team at the CRAL-Observatoire de Lyon), the NASA/IPAC Extragalactic
Database (NED, which is operated by the Jet Propulsion Laboratory,
Caltech, under contract with the National Aeronautics and Space
Administration), and the ADS Abstracts Service.

\bsp

\newpage
\newpage

\newpage

\newpage

\newpage

\newpage

\newpage

\newpage

\newpage

\newpage

\newpage

\newpage

\newpage

\newpage

\newpage

\newpage

\label{lastpage}

\end{document}